\documentclass[cernpreprint, atlasdraft=false, USenglish, texmf, orcidlogo]{atlasdoc}
\usepackage{atlaspackage}
\usepackage{atlasbiblatex}

\usepackage{atlasphysics}
\usepackage{multirow}
\usepackage{lscape}
\usepackage{longtable}
\usepackage{rotating}
\usepackage{float}
\usepackage{bm}
\usepackage{amsmath}
\usepackage{enumitem}

\graphicspath{{logos/}{figures/}}

\usepackage{ANA-BPHY-2023-01-PAPER-defs}

\AtlasJournalRef{Phys. Rev. D 113 (2026) 112015}

\AtlasDOI{10.1103/zhkx-k9zl}

\AtlasTitle{Observation of structures in the $J/\psi +\psi(2S)$ mass spectrum with the ATLAS detector}

\AtlasVersion{3.1}

\AtlasAbstract{%
A search for resonant structures in the $J/\psi +\psi$(2S) mass spectrum is performed using proton--proton collision data at $\sqrt{s}=13$~TeV, corresponding to an integrated luminosity of 140~fb$^{-1}$, recorded by the ATLAS experiment at the LHC\@.
The decay channels of $J/\psi +\psi\text{(2S)} \rightarrow 4\mu$ and $J/\psi +\psi\text{(2S)} \rightarrow 4\mu+2\pi$ are analyzed.
An excess near 6.9 GeV is observed in both channels with a combined significance of 8.9$\sigma$.
No significant signal is observed near 7.2~GeV, and an upper limit on its yield relative to $X$(6900) is provided.
A simultaneous fit with the di-$J/\psi$ channel is carried out under assumptions regarding the resonance interferences, yielding a ratio of the partial decay widths between the $J/\psi +\psi\text{(2S)}$ and di-$J/\psi$ channels of $1.08\pm 0.20_{-0.17}^{+0.40}$ for the resonance near 6.9 GeV.}

\author{The ATLAS Collaboration}

\AtlasRefCode{BPHY-2023-01}

\PreprintIdNumber{CERN-EP-2025-182}

\AtlasJournal{\PRD}
\AtlasCoverEgroupAnalysisTeam{atlas-bphy-2023-01-analysis-team@cern.ch}

\hypersetup{pdftitle={ATLAS document},pdfauthor={The ATLAS Collaboration}}

\begin{document}

\maketitle

%

%
%

\section{Introduction}
\label{sec:int}

The study of tetraquark states ($qq\bar{q}\bar{q}$) can further our understanding of quantum chromodynamics (QCD) in the non-perturbative regime~\cite{Berezhnoy:2011xn,Karliner:2016zzc,Wu:2016vtq}. The topic of all-charm tetraquarks has gained significant interest recently.
The LHCb Collaboration reported the first observation of a narrow resonance near 6.9~\GeV ($X(6900)$) in the di-$J/\psi$ mass spectrum in 2020~\cite{LHCb_TQ}, interpreted as an all-charm tetraquark ($T_{cc\bar{c}\bar{c}}$)~\cite{Iwasaki:1975pv,Chao:1980dv,Wu:2016vtq,Anwar:2017toa,Debastiani:2017msn,Liu:2019zuc,Becchi:2020uvq,Bedolla:2019zwg,ZHU2021115393}. The ATLAS and CMS experiments later confirmed the observation of $X(6900)$, as well as another broad structure around 6.6 GeV~\cite{BPHY-2022-01, CMS-BPH-21-003}.
Recently, CMS reported measurements of the quantum numbers of three all-charm tetraquarks, including a further resonance near 7.1~\GeV, favoring a scenario of tightly bound states~\cite{CMS-BPH-24-002}.
It becomes interesting to search for similar resonances in the $J/\psi$+$\psi$(2S) spectrum and gain more insight into these exotic states. The decay $\psi\text{(2S)}\to\mu\mu$ was investigated by ATLAS in Ref.~\cite{BPHY-2022-01}. In this study, the decay $\psi\text{(2S)}\to J/\psi\pi^+\pi^-\to 2\mu 2\pi$ is also reconstructed, whose total branching fraction is more than twice that of the former.
The analysis uses 140~fb$^{-1}$ of proton--proton ($pp$) collision data collected by the ATLAS experiment at $\sqrt{s}=13$\nobreak\hspace{.24em}~\TeV~ during 2015--2018.

This paper is organized as follows. Section~\ref{sec:det} provides a brief description of the ATLAS detector.  Section~\ref{sec:sim} describes the Monte Carlo (MC) simulation of signal and background processes. The event selection and the definition of the signal regions (SRs) and control regions (CRs) that enter the simultaneous fit are presented in Section~\ref{sec:sel}. The background estimation methods and the additional CRs employed for background determination and validation are detailed in Section~\ref{sec:bkg}. Section~\ref{sec:spectra} shows the $J/\psi$+$\psi$(2S) mass spectra with estimated backgrounds before the fit. The signal modeling and the fit strategy are discussed in Section~\ref{sec:fit_model}, and the fit results are presented in Section~\ref{sec:res}. The evaluation of systematic uncertainties is described
in Section~\ref{sec:sys}. Finally, the conclusions are summarized in Section~\ref{sec:sum}.

\section{The ATLAS detector}
\label{sec:det}
The ATLAS experiment~\cite{PERF-2007-01} is a multipurpose particle detector with a forward--backward symmetric cylindrical geometry and nearly $4\pi$ coverage in solid angle.\footnote{\AtlasCoordFootnote}
It consists of an inner tracking detector (ID) surrounded by a thin superconducting solenoid providing a 2T (Tesla) axial magnetic field, sampling electromagnetic and hadronic calorimeters and a muon spectrometer (MS).
The muon and tracking systems play a crucial role in reconstructing signal events.
The ID covers the pseudorapidity range $|\eta|<2.5$ and consists of a silicon pixel detector, a silicon microstrip detector, and a transition radiation tracker.
The MS surrounds the calorimeters and is based on three large superconducting air-core toroids with eight coils
each. It includes a system of tracking chambers up to $|\eta| = 2.7$ and fast detectors for triggering up to $|\eta| = 2.4$.  A two-level trigger system is used to select events for offline analysis~\cite{TRIG-2019-04}. A software suite~\cite{SOFT-2022-02} is used in data simulation, in the reconstruction and analysis of real and simulated data, in detector operations, and in the trigger and data acquisition systems of the experiment. Events are selected for analysis only if all detector systems are in good operating condition~\cite{DAPR-2018-01}.


\section{Simulation of signal and background processes}
\label{sec:sim}

MC simulated samples are used to study the mass resolution and signal efficiency and to estimate some of the background processes.
The main backgrounds are prompt di-charmonium ($J/\psi$ or $\psi\text{(2S)}$) production via single~\cite{Qiao:2002rh,Lansberg1} and double~\cite{Calucci2,Kom} parton scattering (SPS and DPS), Nonprompt charmonium production from $b$-hadron (via $b\bar{b}$ production) decays where the charmonium vertex is displaced from the $pp$ collision vertex, prompt single charmonium production, and non-resonant di-muon production. The last two enter the analysis due to tracks misidentified as muons and are collectively categorized as Others.
\Pythia 8.244~\cite{Sjostrand:2014zea} was used to generate the other three backgrounds. Both the color-singlet and color-octet intermediate states were included for $J/\psi$ and $\psi$(2S).
The signal was generated with \Pythia 8.306~\cite{Bierlich:2022pfr}.
The A14 set of tuned parameters~\cite{ATL-PHYS-PUB-2014-021} and the NNPDF23LO~\cite{Ball:2014uwa} parton distribution function were used. Final-state radiation in particle decays was simulated by PHOTOS~\cite{Davidson:2010ew}. Exclusive decays of $J/\psi \to 2\mu$ and $\psi(2S) \to 2\mu$ or $\psi(2S) \to J/\psi \pi^+\pi^- \to 2\mu 2\pi$ are enforced.
The decay products are evenly distributed in the final state phase space. Studies based on MC simulation show that the ratio of signal efficiencies between the two $\psi(2S)$ decay modes exhibits only a weak dependence on this assumption, which is relevant for the ratio of signal yields in the different channels.



\section{Event selection}
\label{sec:sel}

The events are collected by di-muon and tri-muon trigger paths with at least one di-muon pair whose mass is consistent with $J/\psi$ or $\psi(\text{2S})$, and with trigger muon $p_{\text{T}}$ thresholds tuned for different data-taking periods~\cite{TRIG-2018-01}.
The trigger efficiency relative to offline requirements for both signals $X\text{(6900)}\to J/\psi+\psi\text{(2S)}\to 4\mu(+2\pi)$ is about $65\%$.
Muons are reconstructed by matching tracks found in the ID and the MS. They are required to satisfy the \textit{loose} identification criteria~\cite{MUON-2018-03}.
In all channels, events are required to contain at least four muons forming two opposite-charge pairs. In the $4\mu+2\pi$ channel, two additional opposite-sign (OS) charged tracks fulfilling the \textit{loose} criteria~\cite{ATL-PHYS-PUB-2015-051,PERF-2015-08} are required. The ID tracks of four muons (plus two pions for $4\mu+2\pi$) are fitted to a common vertex with the mass of each charmonium constrained~\cite{ATL-PHYS-2003-031} to its world average value~\cite{Zyla:2020zbs}.
Additional selection requirements are imposed on variables such as the muon momentum, vertex fit quality, and the projected transverse distances between the primary vertex~\cite{ATL-PHYS-PUB-2015-026} and the candidate vertex. The event selection criteria for the SRs and CRs are summarised in Table~\ref{tab:selection}.
The definitions of the variables in the table are listed below.

\begin{enumerate}[label=(\roman*)]
\item The definitions of $m_{J/\psi}$, $m_{\psi(2S)}$. The charmonium invariant mass before the vertex fit.
\item The definitions of $m_{4\mu}$, $m_{4\mu+2\pi}$. The $4\mu$ or $4\mu+2\pi$ vertex mass after the vertex fit.
\item The definitions of $\chi^2_{4\mu}/N$, $\chi^2_{4\mu+2\pi}/N$. The $4\mu$ or $4\mu+2\pi$ vertex fit quality $\chi^2$ divided by the number of degrees of freedom $N$. The vertex candidate with the lowest $\chi^2/N$ is selected in events with more than one candidate meeting the criteria on $\chi^2/N$.
\item The definitions of $L_{xy}^{4\mu}$, $L_{xy}^{4\mu+2\pi}$, $L_{xy}^{\text{charm}}$. The distance in the transverse plane from the primary vertex to the $4\mu$, $4\mu+2\pi$ vertices and charmonium subvertex, projected onto the direction of the vertex candidate's transverse momentum, respectively.
The charmonium subvertex is obtained by revertexing the tracks from its decay with the subvertex mass constrained to the charmonium's mass~\cite{Zyla:2020zbs}.
The primary vertex is the $pp$ collision vertex. If there are multiple primary vertices in the event due to pileup~\cite{ATL-PHYS-PUB-2014-001}, the one closest to the candidate vertex in $z$ is chosen. These quantities can be negative due to the resolution effect.
\end{enumerate}

\begin{table*}[htb]
\caption{Event selection criteria for the signal and control regions in the $4\mu$ and $4\mu+2\pi$ channels.}
\begin{center}
\small
\begin{tabular}{c|c|c|c}
\toprule
\multicolumn{2}{c|}{$4\mu$ channel} & \multicolumn{2}{c}{$4\mu+2\pi$ channel} \\ \midrule
SR & CR & SR & CR \\ \midrule
\multicolumn{4}{c}{Di-muon or tri-muon triggers, oppositely charged muons from each charmonium,} \\
\multicolumn{4}{c}{\textit{Loose}  muons, $p_{\text{T1,2,3,4}}>4,4,3,3$~\GeV~and $|\eta_{1,2,3,4}|<2.5$ for the four muons,} \\
\multicolumn{4}{c}{$m_{J/\psi}\in [2.94,3.25]$~\GeV, $m_{\psi(2S)}\in [3.56,3.80]$~\GeV} \\ \midrule

\multicolumn{2}{c|}{---} & \multicolumn{2}{c}{Two \textit{loose} OS ID tracks with $p_{\text{T}}>0.5$~\GeV } \\
\multicolumn{2}{c|}{} & \multicolumn{2}{c}{for pions, BDT requirement} \\ \midrule

\multicolumn{2}{c|}{$\chi^2_{4\mu}/N < 3$, $|L_{xy}^{4\mu}| < 0.2$ mm,} & \multicolumn{2}{c}{$\chi^2_{4\mu+2\pi}/N < 3$, $|L_{xy}^{4\mu+2\pi}|< 0.2$ mm,} \\
\multicolumn{2}{c|}{$|L_{xy}^{\text{charm}}| < 0.3$ mm, $m_{4\mu}<11$~\GeV} & \multicolumn{2}{c}{$|L_{xy}^{\text{charm}}| < 0.3$ mm, $m_{4\mu+2\pi}<11$~\GeV} \\ \midrule
$\Delta R (J/\psi, \psi(2S))<0.25$ & $\Delta R (J/\psi, \psi(2S))\ge 0.25$ & $\Delta R (J/\psi, \psi(2S))<0.25$ & $\Delta R (J/\psi, \psi(2S))\ge 0.25$ \\

\bottomrule

\end{tabular}
\label{tab:selection}
\end{center}
\end{table*}

For the $4\mu+2\pi$ channel, an additional set of regions is defined to constrain the feed-up background, and a boosted decision tree (BDT) is employed to further suppress this background (details in Section~\ref{sec:feedup}). These regions follow the same selection requirements as those listed in Table~\ref{tab:selection}, except that the two pion candidates are required to have same-sign (SS) charges. Analogous to the definitions of the SR and CR, the regions are classified as the SS SR-like and SS CR-like regions based on the same $\Delta R$ criterion.

The overall analysis strategy follows that of~\cite{BPHY-2022-01}, in which the $\Delta R$ between the two charmonia is used to define the SRs and the CRs, as shown in the bottom row of Table~\ref{tab:selection}.
Although this variable has some correlation with the di-charmonium mass, imposing a requirement on it provides three advantages: it reduces the background in the SRs, constrains the background normalization, and validates the SPS background shape.
The agreement of the data with the predicted background shape in the CRs allows to validate the setting of the \Pythia parameter \texttt{pT0timesMPI}~\cite{Sjostrand:2014zea} that was tuned on data~\cite{BPHY-2022-01} (more details on the \Pythia parameters are given in Section~\ref{sec:common_backgrounds}). This approach ensures that the background estimation is robust before examining the data events in the SRs.

%

%
%
%


\section{Background estimation}
\label{sec:bkg}

\subsection{Common backgrounds}
\label{sec:common_backgrounds}

The backgrounds common to the $4\mu$ and $4\mu+2\pi$ channels are estimated sequentially in the order of Others, Nonprompt, DPS, and SPS.
The Others background, consisting of events with at most one real charmonium, is estimated by a data-driven method based on the charmonium mass sideband events.
The Nonprompt charmonium background is estimated with MC simulation, which is corrected and normalized in a Nonprompt CR defined by primarily reversing the $L_{xy}^{\text{charm}}$ requirement.
The SPS and DPS di-charmonium backgrounds are also estimated using MC simulation.
Dedicated CRs are defined to constrain the SPS and DPS background contributions.
The SPS (DPS) CR is defined by requiring the di-charmonium mass to fall within [8, 12] ([14, 25])~\GeV, where the corresponding background contribution is dominant.
The modeling of the SPS background depends on the \Pythia parameter \texttt{pT0timesMPI}. This parameter is part of the \texttt{SuppressSmallPT} settings~\cite{Sjostrand:2014zea} used to regularize the small-$p_{\text{T}}$ behavior of $2\to2$ scattering processes, and it has a significant impact on the shape of the di-charmonium mass spectrum predicted by the simulation.
The nominal value of \texttt{pT0timesMPI} was determined by comparing the di-charmonium mass distributions in data and simulation in the SPS CR of the di-$J/\psi$ channel in Ref.~\cite{BPHY-2022-01}.

The event selection criteria for various CRs used for background estimation are listed in Table~\ref{tab:bkgCR}. The Others, Nonprompt and DPS CRs are orthogonal to the SRs and CRs listed in Table~\ref{tab:selection}, whereas the SPS CR has some overlap with them. This overlap occurs for the sideband events in the SRs with $m_{4\mu(+2\pi)}>8$~\GeV, where the signal contribution is negligible.

\begin{table*}[htb]
\caption{Requirements for the background CRs that differ from the criteria as listed in Table~\ref{tab:selection}.}
\begin{center}
\begin{tabular}{c|c|c|c}
\toprule
Others & Nonprompt & DPS & SPS \\ \midrule
$m_{J/\psi}\in [2.60, 2.76]$~\GeV or $m_{J/\psi}$ & $L^{\text{charm}}_{xy}>0.3$ mm for & No $\Delta R$ or BDT & No $\Delta R$ or BDT \\
$\in [3.34, 3.50]$~\GeV for either $J/\psi$; & either charmonium, & requirements, & requirements, \\
$m_{\psi\text{(2S)}}\in [3.35, 3.48]$~\GeV or $m_{\psi\text{(2S)}}$ & $\chi^2_{4\mu(+2\pi)}/N<15$, & $m_{4\mu(+2\pi)}$ & $m_{4\mu(+2\pi)}$ \\
$\in [3.88, 4.10]$~\GeV for $\psi$(2S) & $m_{4\mu(+2\pi)}<25$~\GeV & $\in [14, 25]$~\GeV & $\in [8, 12]$~\GeV \\
\bottomrule
\end{tabular}
\label{tab:bkgCR}
\end{center}
\end{table*}

The background CRs are used both to derive kinematic corrections and to normalize the relevant background contributions. Several kinematic variables are corrected by assigning event weights to the MC events, ensuring agreement with the data in the background CRs. These weights also improve the data–MC agreement for other variables whose distributions are not directly corrected. The corrected variables are listed in Table~\ref{tab:bkgCorr}.
The correction weights are derived as polynomial functions of up to third order, obtained from the ratio of data to MC distributions. The value of each correction function is multiplied with the event weight. Due to correlations among the corrected variables, adjusting one
variable can degrade the data–MC agreement for others; therefore, the correction procedure is iterated several times. The iterations are terminated once the parameters of the correction functions from successive iterations converge within their respective uncertainties. Once derived from
the background CRs, the corrections are applied to all regions.

\begin{table*}[htb]
\caption{List of variables corrected in the different background CRs.}
\begin{center}
\begin{tabular}{c|c|c|c}
\toprule
Others & Nonprompt & DPS & SPS \\ \midrule
$4^\text{th}$ muon $p_{\text{T}}$, &  & di-charmonium $p_{\text{T}}$, & di-charmonium $p_{\text{T}}$, \\
charmonium $p_{\text{T}}$, & $4^\text{th}$ muon $p_{\text{T}}$ & $\Delta\eta$ between charmonia, & $\Delta\eta$ and $\Delta\phi$ between charmonia, \\
$m_{4\mu(+2\pi)}$ &  & $4^\text{th}$ muon $p_{\text{T}}$, $2^\text{nd}$ pion $p_{\text{T}}$ & $4^\text{th}$ muon $p_{\text{T}}$, $2^\text{nd}$ pion $p_{\text{T}}$ \\
\bottomrule
\end{tabular}
\label{tab:bkgCorr}
\end{center}
\end{table*}

Because the contributions of Nonprompt and DPS backgrounds are very small in the SR, it is crucial to control the di-charmonium mass shape for the SPS and Others background components in this region. In the $J/\psi+\psi\text{(2S)}\to 4\mu$ channel, the Others contribution is also important due to the lower signal purity under the $\psi\text{(2S)}$ mass peak compared to the $J/\psi$.
The Others background is modeled from data itself by requiring at least one charmonium candidate to contain a track that fails the muon identification criteria. Charmonium mass sideband events are used to determine the normalization and corrections for this background.
Its modeling is validated in the Nonprompt region defined in Table~\ref{tab:bkgCR}, where the Others contribution dominates and the Nonprompt background is negligible. Any mass shape difference between background and data is assigned as an additional systematic uncertainty (``non-closure'').

In the following sections, an additional global CR is used for validation purposes. This region is defined by requiring the di-charmonium mass to lie within $[8,25]$~GeV, while all other selection requirements are identical to those used for the SPS and DPS CRs.

\subsection{Feed-up background}
\label{sec:feedup}

\begin{figure}[htbp]
\centering
\includegraphics[width=0.45\textwidth]{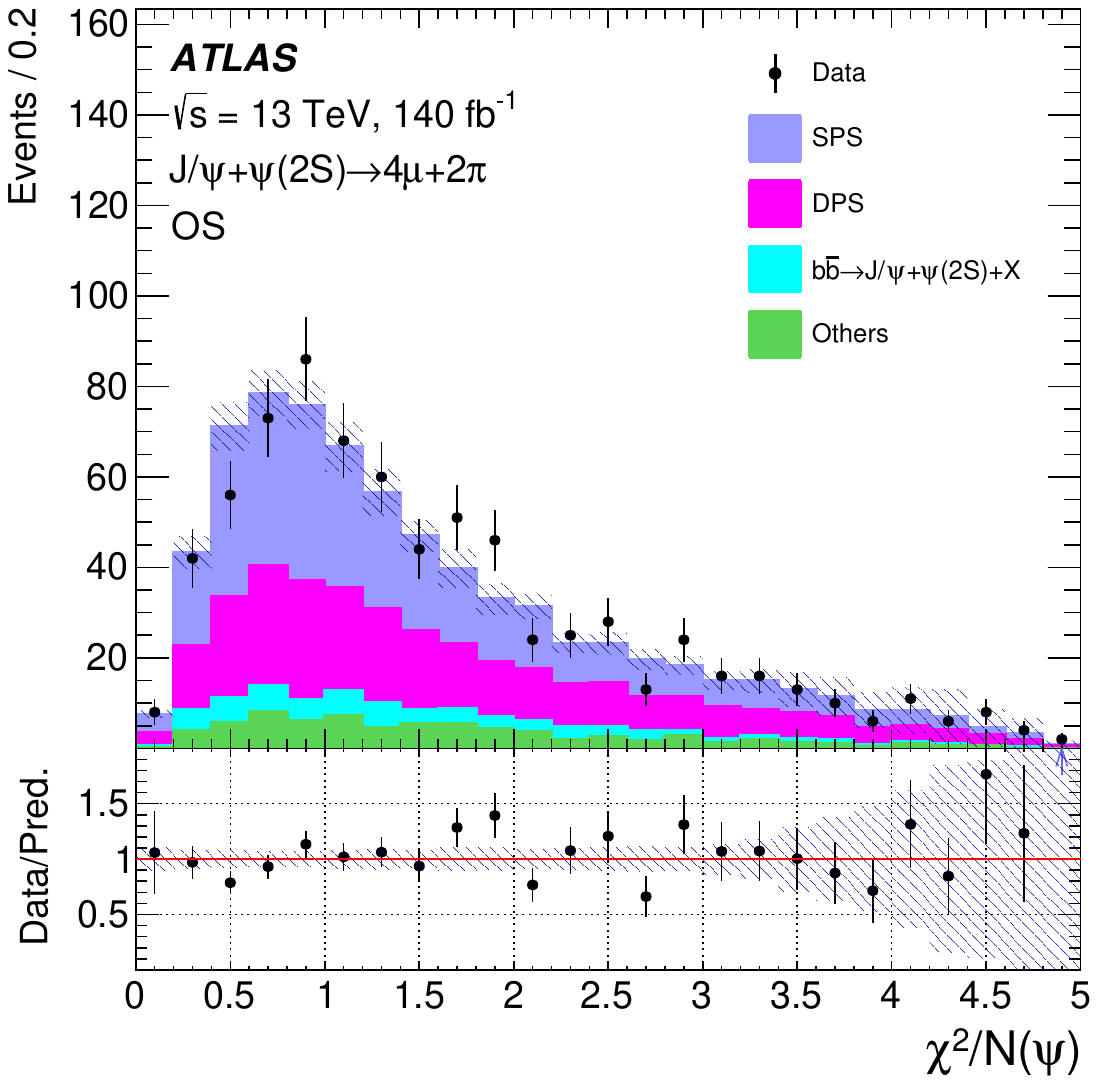}
\includegraphics[width=0.45\textwidth]{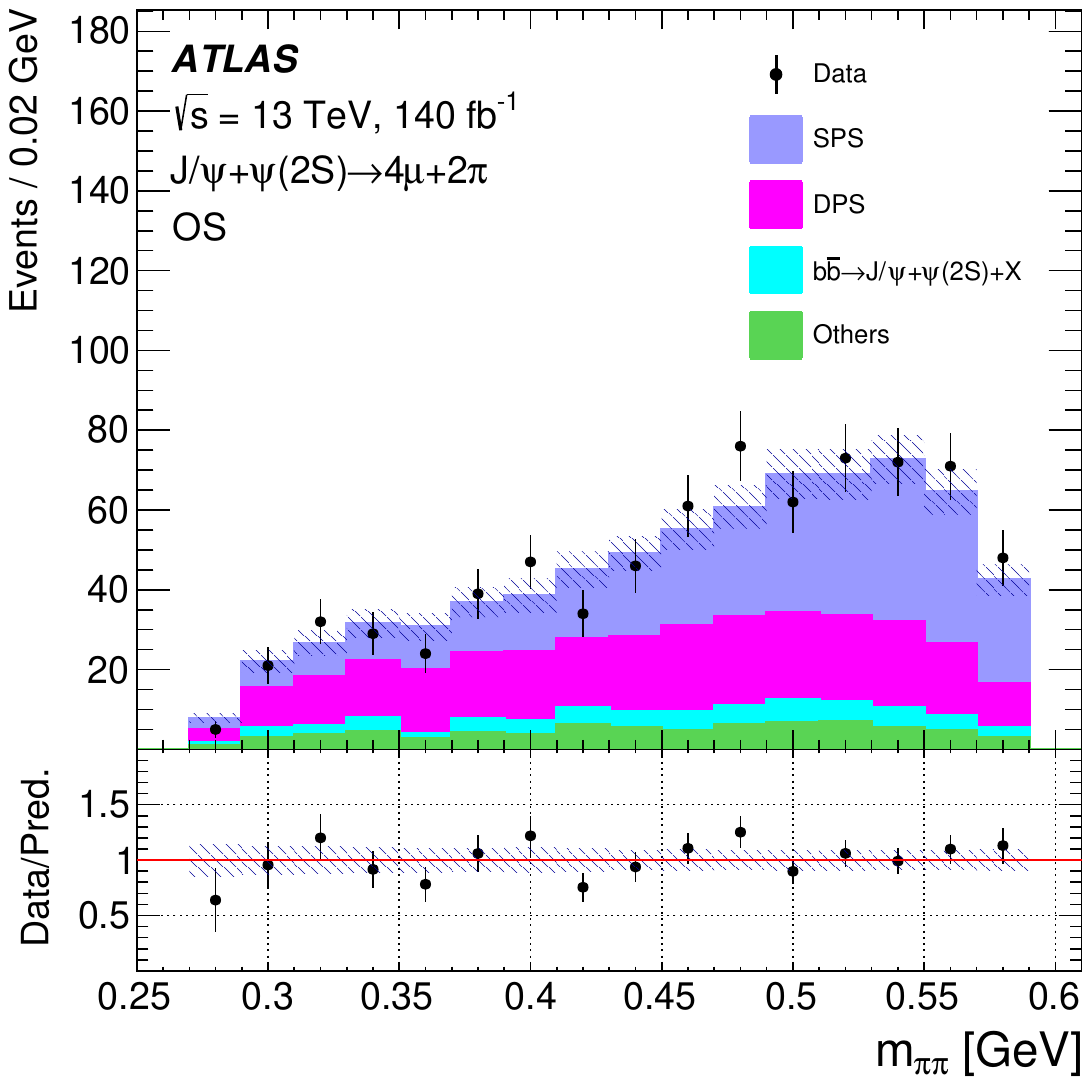}
\includegraphics[width=0.45\textwidth]{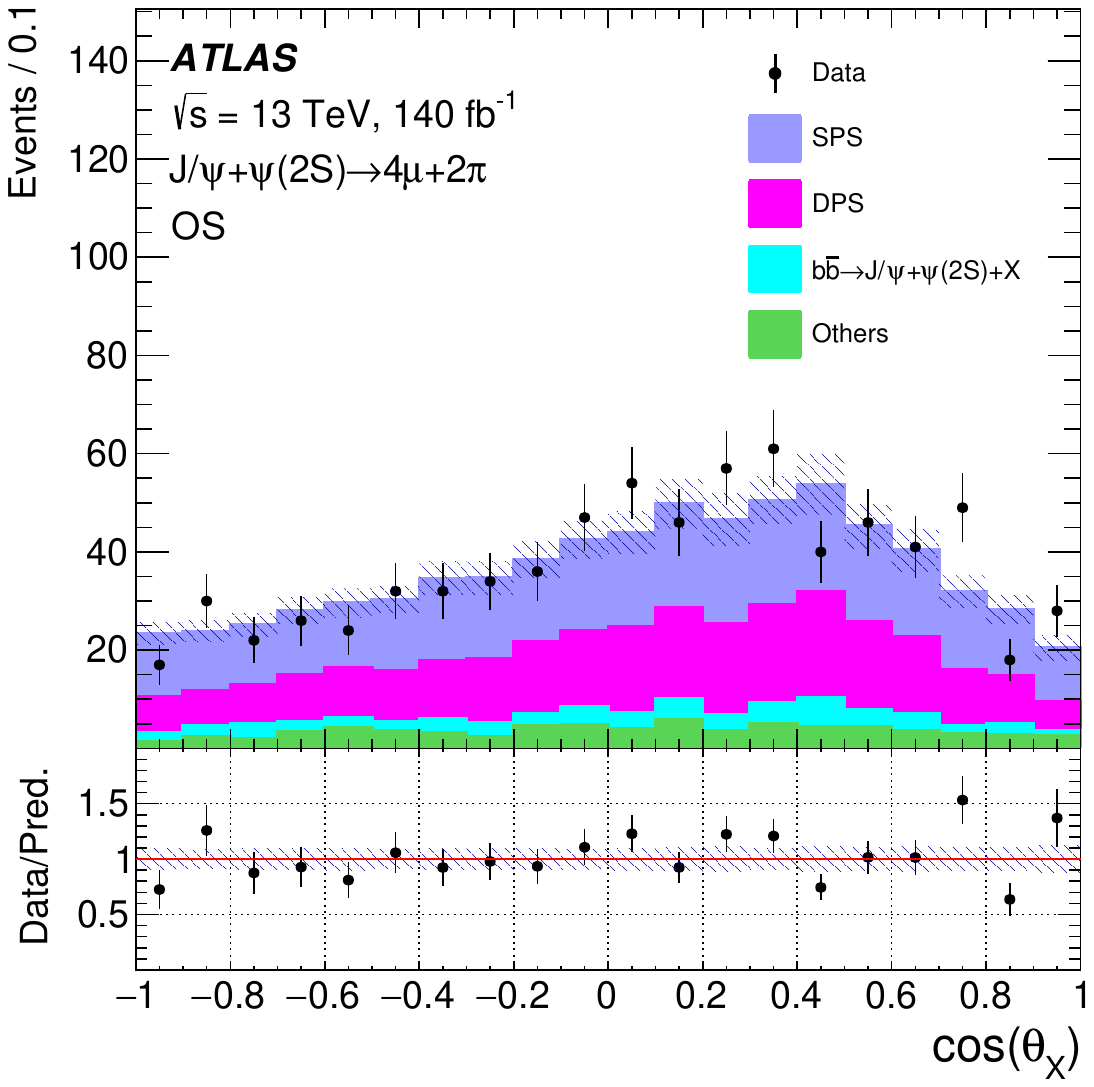}
\includegraphics[width=0.45\textwidth]{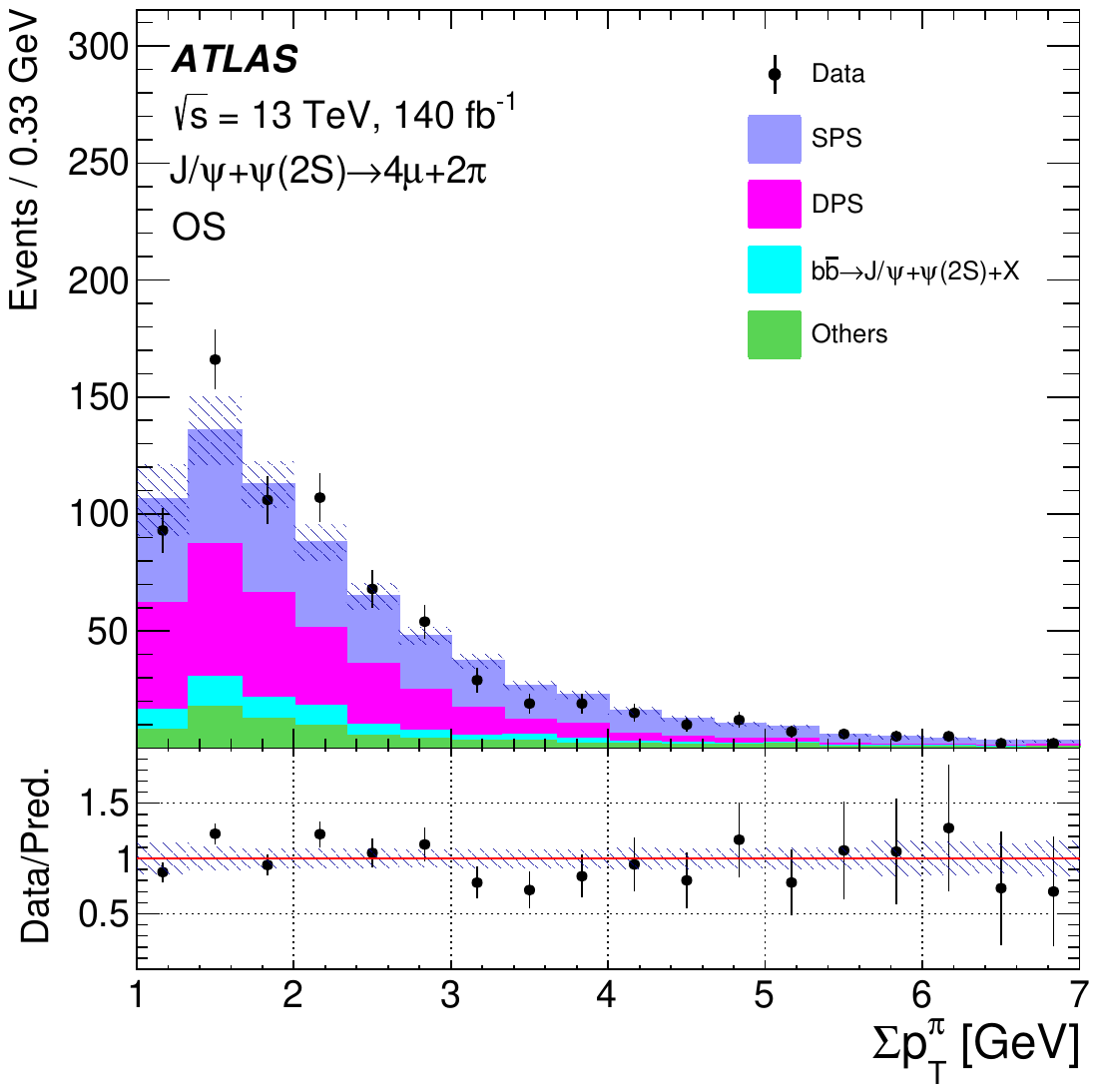}
\includegraphics[width=0.45\textwidth]{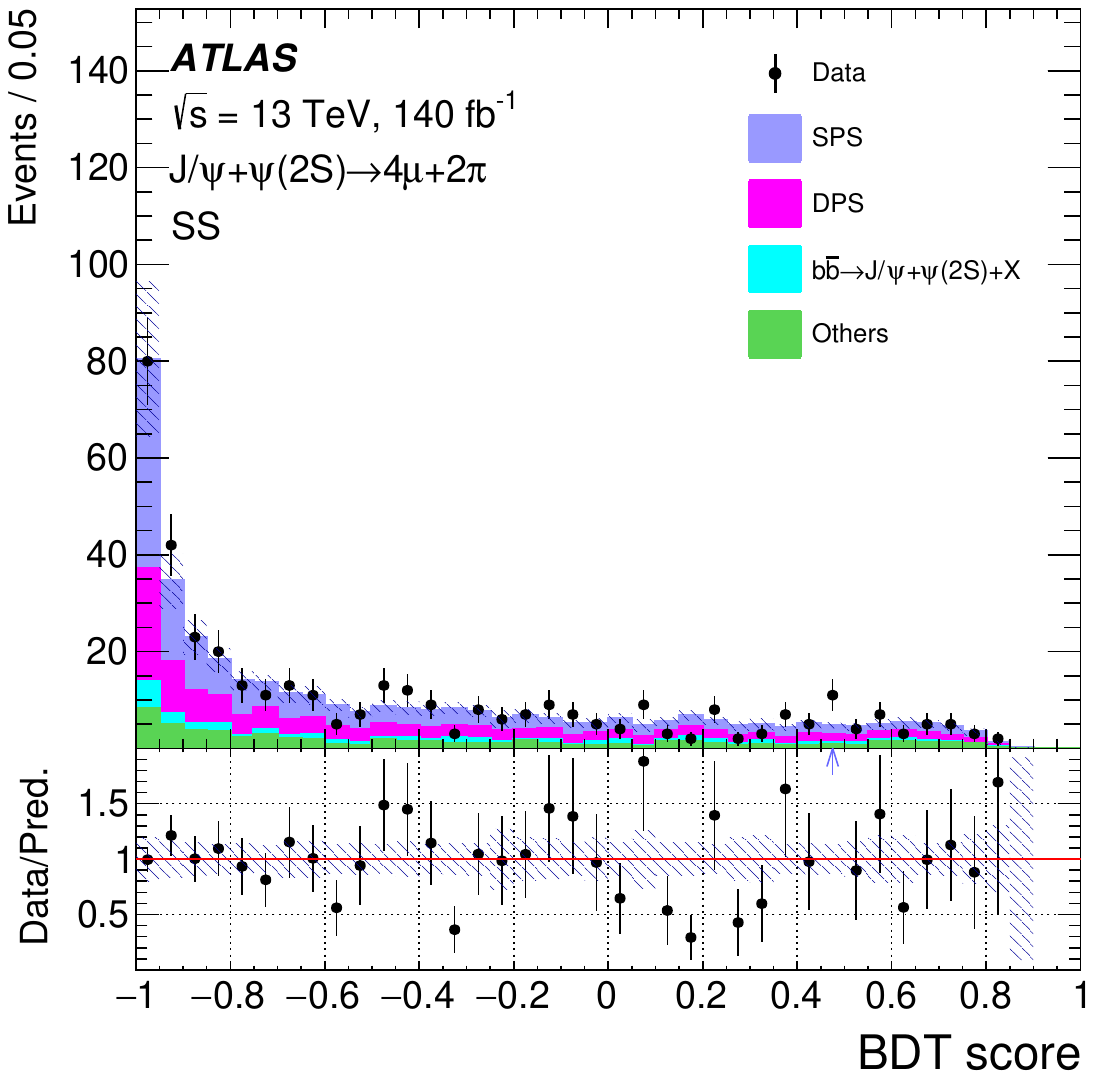}
\includegraphics[width=0.45\textwidth]{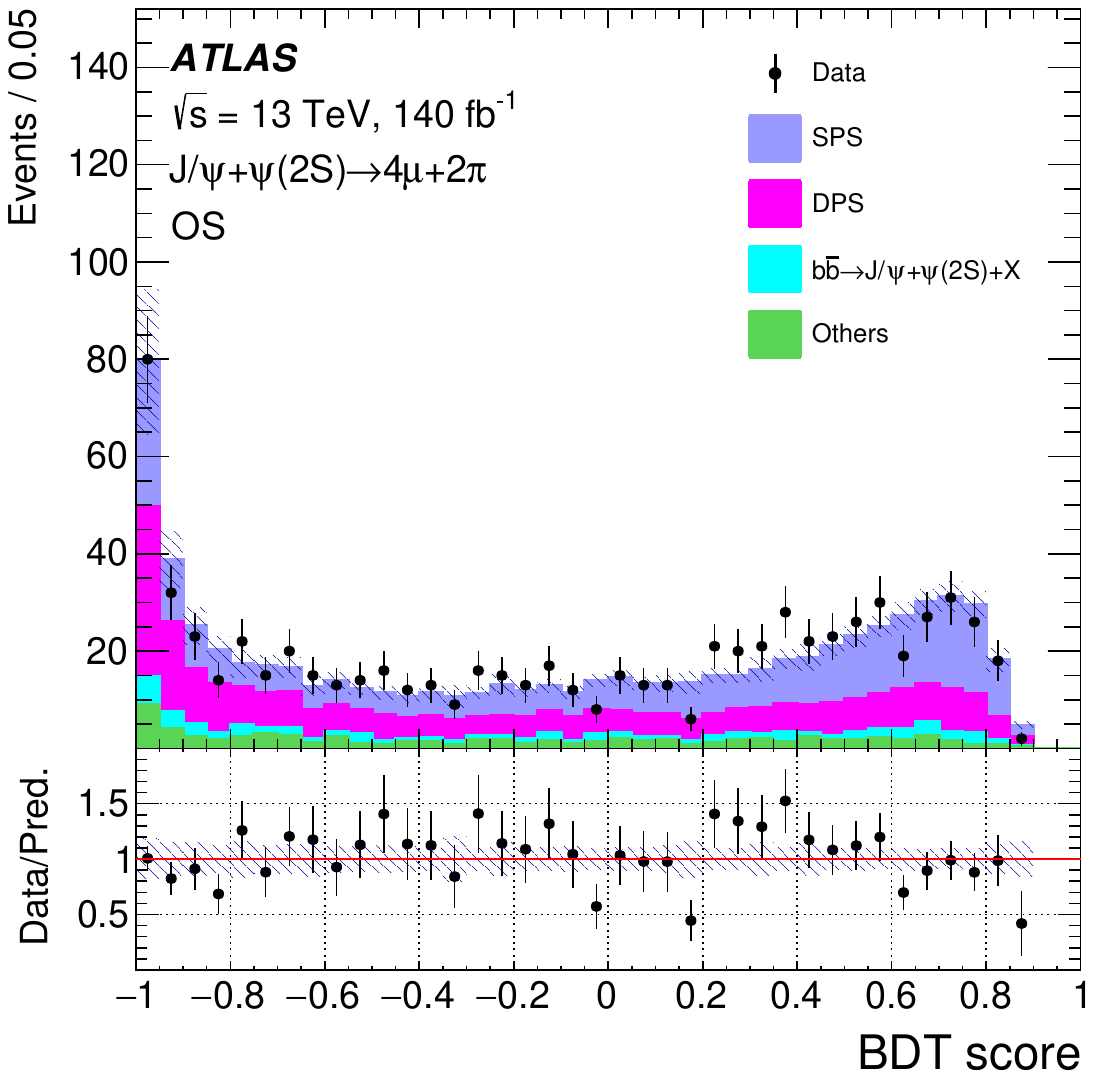}
\caption{ Distributions of the input variables (top and middle) and the BDT scores for SS (bottom left) and OS (bottom right) events in the global CR of the $4\mu+2\pi$ channel. $b\bar{b}\to \text{charmonia}+X$ represents the Nonprompt background. The error bars (shaded areas) represent uncertainties in the data (background predictions) in each bin. Arrows in the lower panels indicate bins in which the ratio of data to the background prediction falls outside the displayed range.}
\label{fig:BDT}
\end{figure}

The $J/\psi+\psi\text{(2S)}\to 4\mu+2\pi$ channel receives a feed-up background from the di-$J/\psi \to 4\mu $ channel, where four muons are combined with two random charged tracks reconstructed under the pion hypothesis.
A BDT with gradient boost~\cite{TMVA} is trained to suppress backgrounds in which the pions do not originate from genuine $\psi$(2S) decays.
In the BDT training, signal is defined by MC events containing correctly reconstructed true $\psi\text{(2S)}\to J/\psi+2\pi$ decays, while background corresponds to those in which the pions are randomly assigned.
The BDT input variables include $\chi^2/N$ of the $\psi$(2S) subvertex, the invariant mass of the two pions ($m_{\pi\pi}$), the cosine of the angle between the $J/\psi$ direction in the $\psi$(2S) rest frame and the $\psi$(2S) direction in the lab frame ($\cos\theta_X$), and the scalar sum of two pions' transverse momenta ($\sum p_{\text{T}}^{\pi}$).
Because the pions from $\psi$(2S) decays follow a phase space distribution in the MC, the $m_{\pi\pi}$ spectrum is reweighted according to the Voloshin-Zakharov model~\cite{Voloshin:1980zf} with the model parameter $\lambda$ taken from Ref.~\cite{BPHY-2015-03}.
Distributions of these input variables and the BDT scores in the global CR are shown in Figure~\ref{fig:BDT}. %
Events with $\psi$(2S) candidates containing random pions, including two SS pions, tend to have low BDT scores. The score is required to exceed $-0.3$ in the SR. The separation power~\cite{TMVA} of the input variables is listed in Table~\ref{tab:sepPower}, where higher values indicate a greater ability to discriminate between signal and background events. With this selection, the feed-up background is reduced by approximately 65\% with the BDT, while about 92\% of the signal is retained.

\begin{figure}[htbp]
\centering
\includegraphics[width=0.45\textwidth]{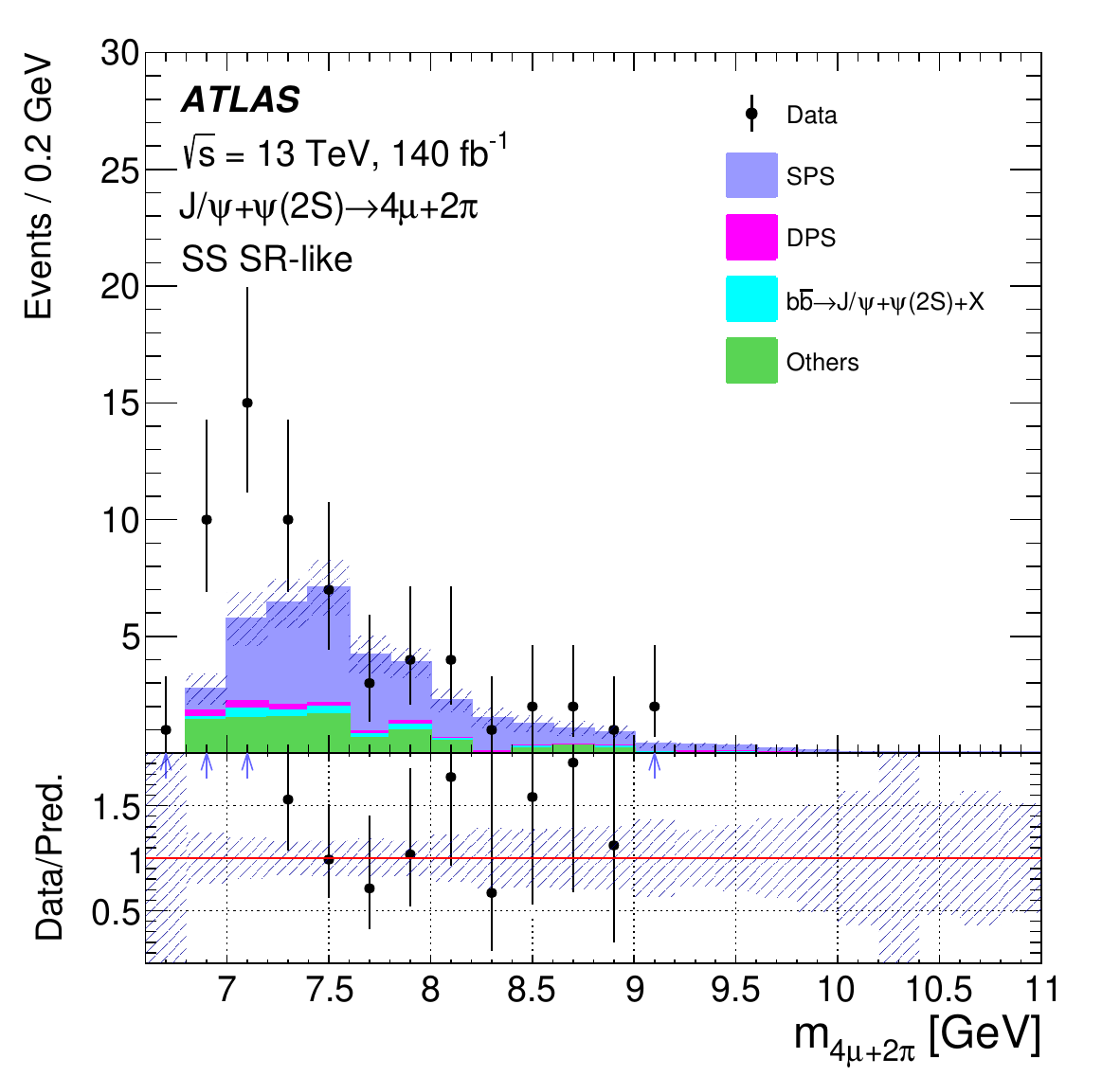}
\caption{The $J/\psi+\psi$(2S) mass spectrum in the SS SR-like region of the $4\mu+2\pi$ channel. The excess of data over the total background arises from the feed-up contribution. }
\label{fig:SS}
\end{figure}

\begin{table*}[htb]
\caption{The separation power of the four BDT input variables.}
\begin{center}
\begin{tabular}{c|c|c|c}
\toprule
$m_{\pi\pi}$ & $\sum p_{\text{T}}^{\pi}$ & $\cos\theta_X$ & $\chi^2/N$ of $\psi$(2S) \\ \midrule
$0.252$ & $0.178$ & $0.106$ & $0.097$ \\
\bottomrule
\end{tabular}
\label{tab:sepPower}
\end{center}
\end{table*}
Because the kinematics of the feed-up background are similar between the OS and SS configurations, its yield in the SR and CR after the BDT selection is constrained by the SS SR-like and SS CR-like regions in the simultaneous fit described in Section~\ref{sec:fit_model}. The corresponding mass shape is obtained from MC simulation. The di-charmonium mass spectrum in the SS SR-like region with $\Delta R < 0.25$ is shown in Figure~\ref{fig:SS}.

\section{$J/\psi$+$\psi$(2S) mass spectra}
\label{sec:spectra}

\begin{figure}[htbp]
\centering
\includegraphics[width=0.45\textwidth]{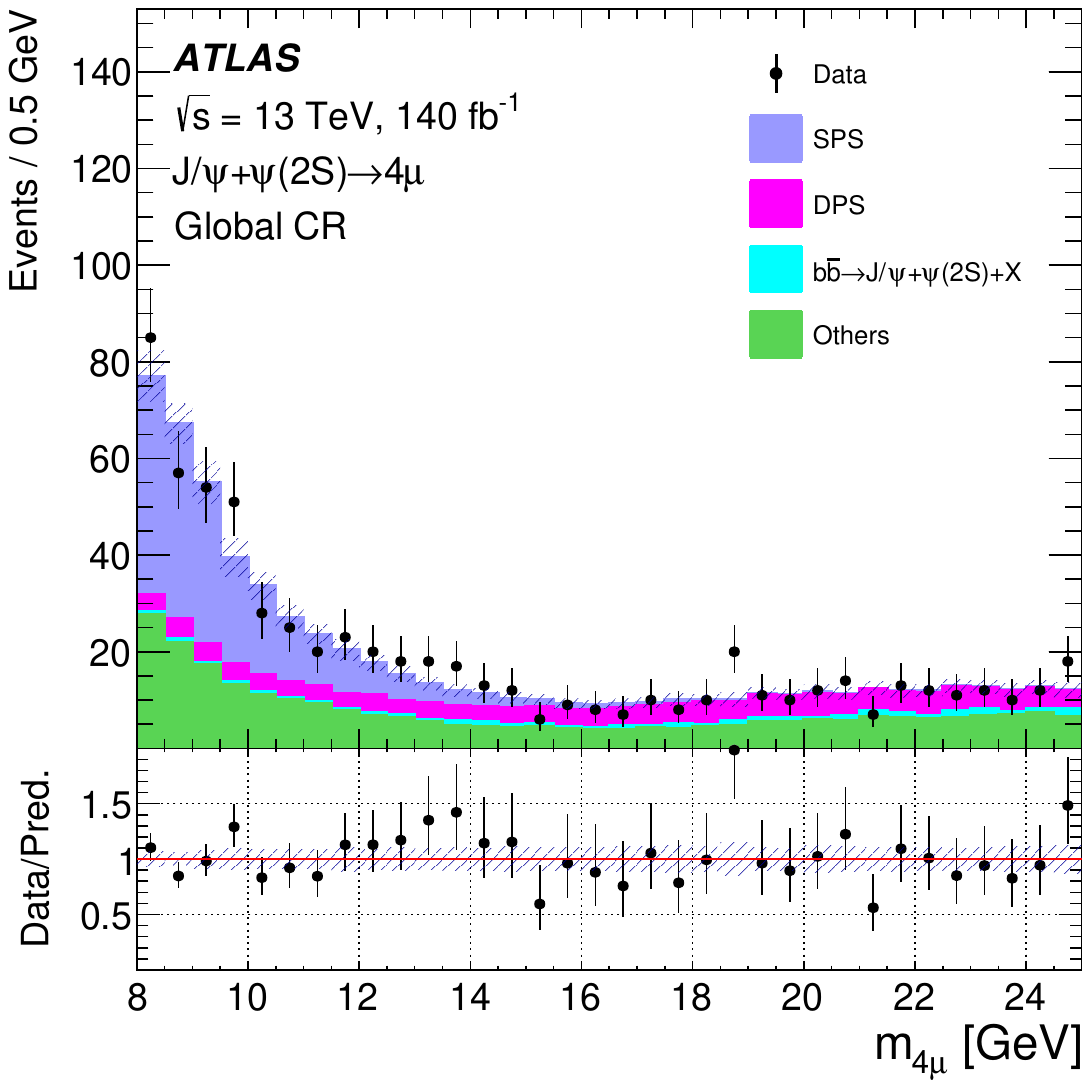}
\includegraphics[width=0.45\textwidth]{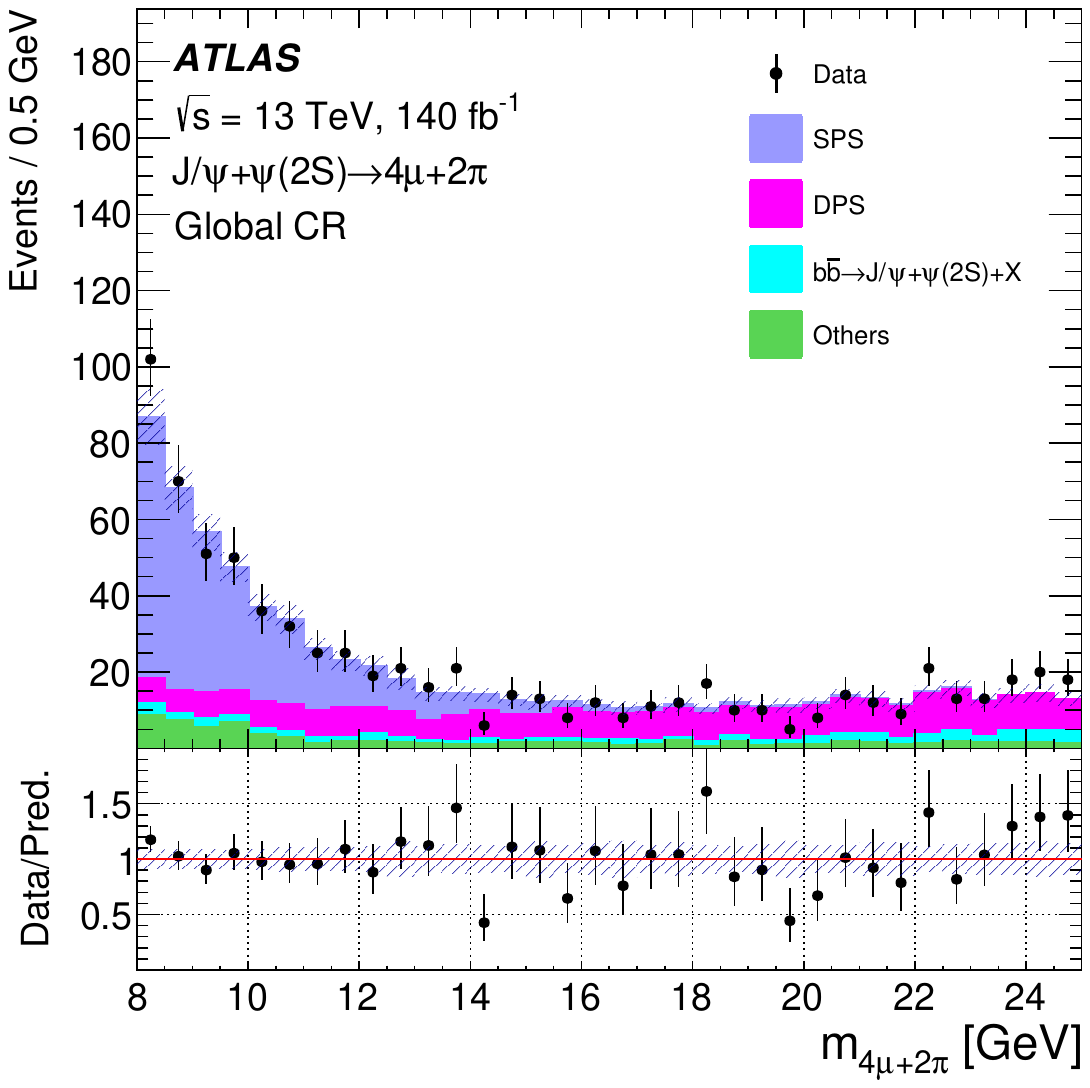}
\caption{The $J/\psi$+$\psi$(2S) mass spectra in the global CR for the $4\mu$ (left) and $4\mu+2\pi$ (right) channels. $b\bar{b}\to \text{charmonia}+X$ represents the Nonprompt background. The error bars (shaded areas) represent uncertainties in the data (background predictions) in each bin. Arrows in the lower panels indicate bins in which the ratio of data to the background prediction falls outside the displayed range.}
\label{fig:VR}
\end{figure}

To validate the background estimation, the $J/\psi$+$\psi$(2S) mass spectra in the global CR are examined for the $4\mu$ and $4\mu+2\pi$ channels. As shown in Figure~\ref{fig:VR}, good agreement between the data and the prediction is observed.

\begin{figure}[htbp]
\centering
\includegraphics[width=0.45\textwidth]{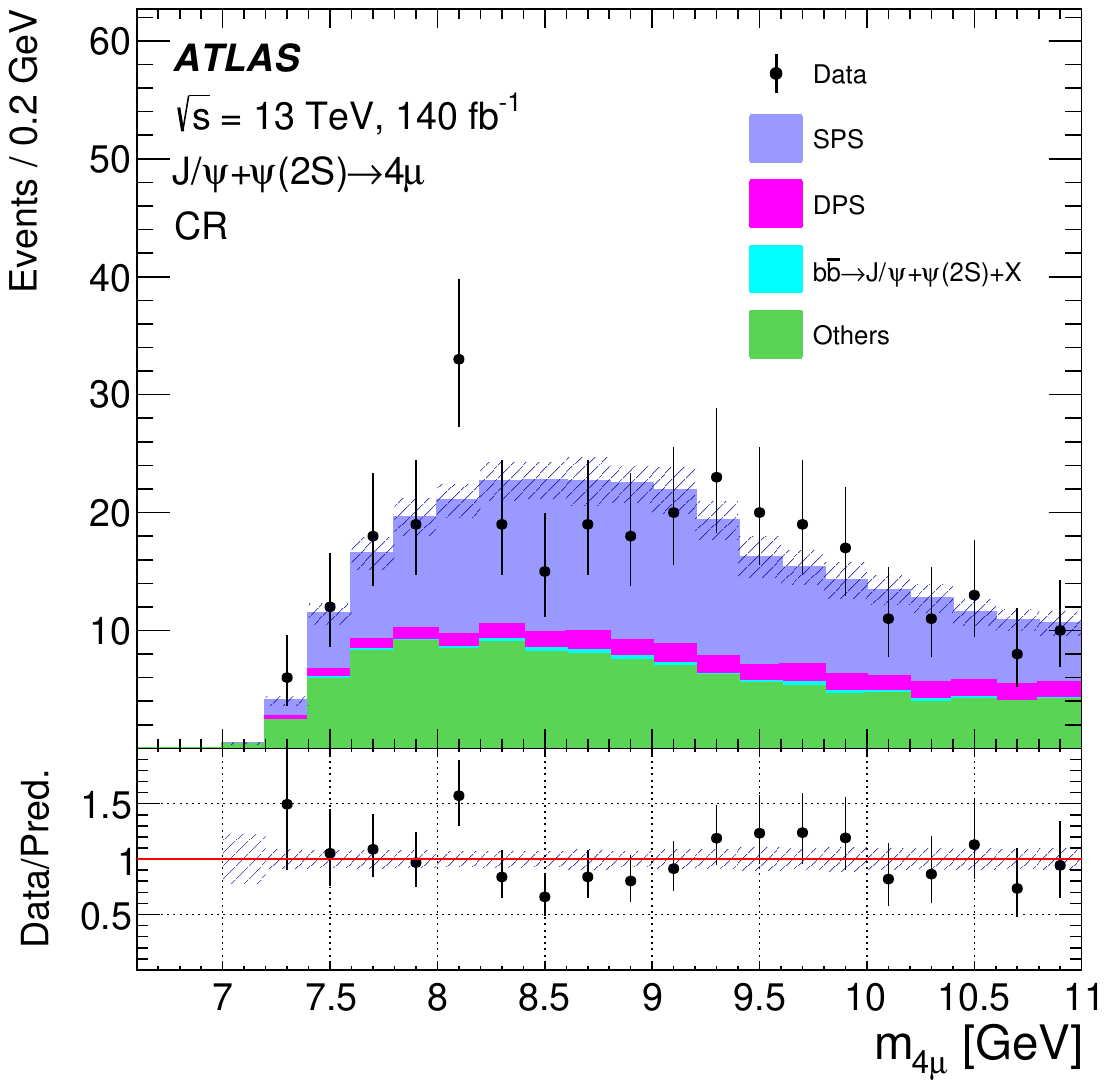}
\includegraphics[width=0.45\textwidth]{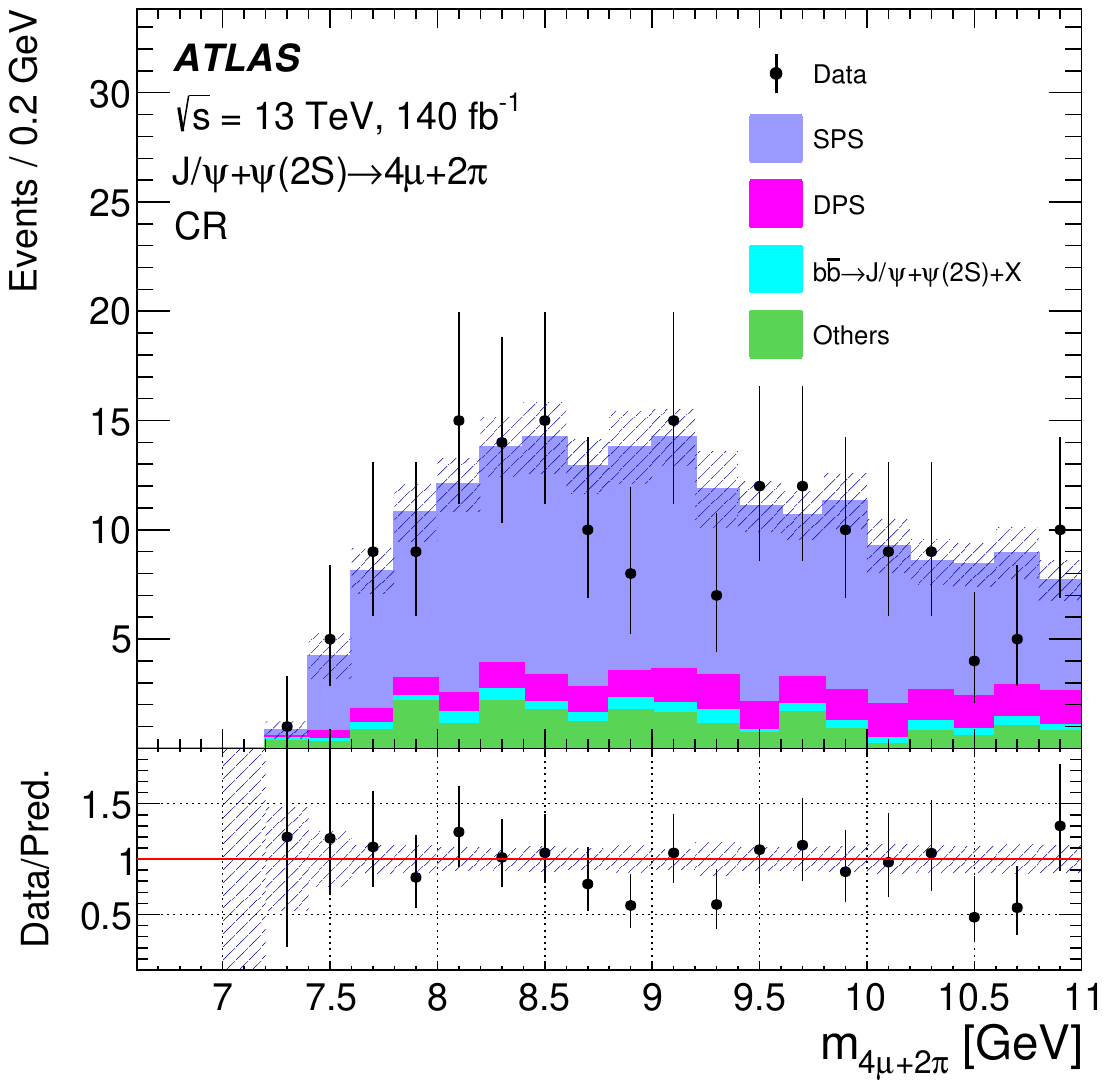}
\includegraphics[width=0.45\textwidth]{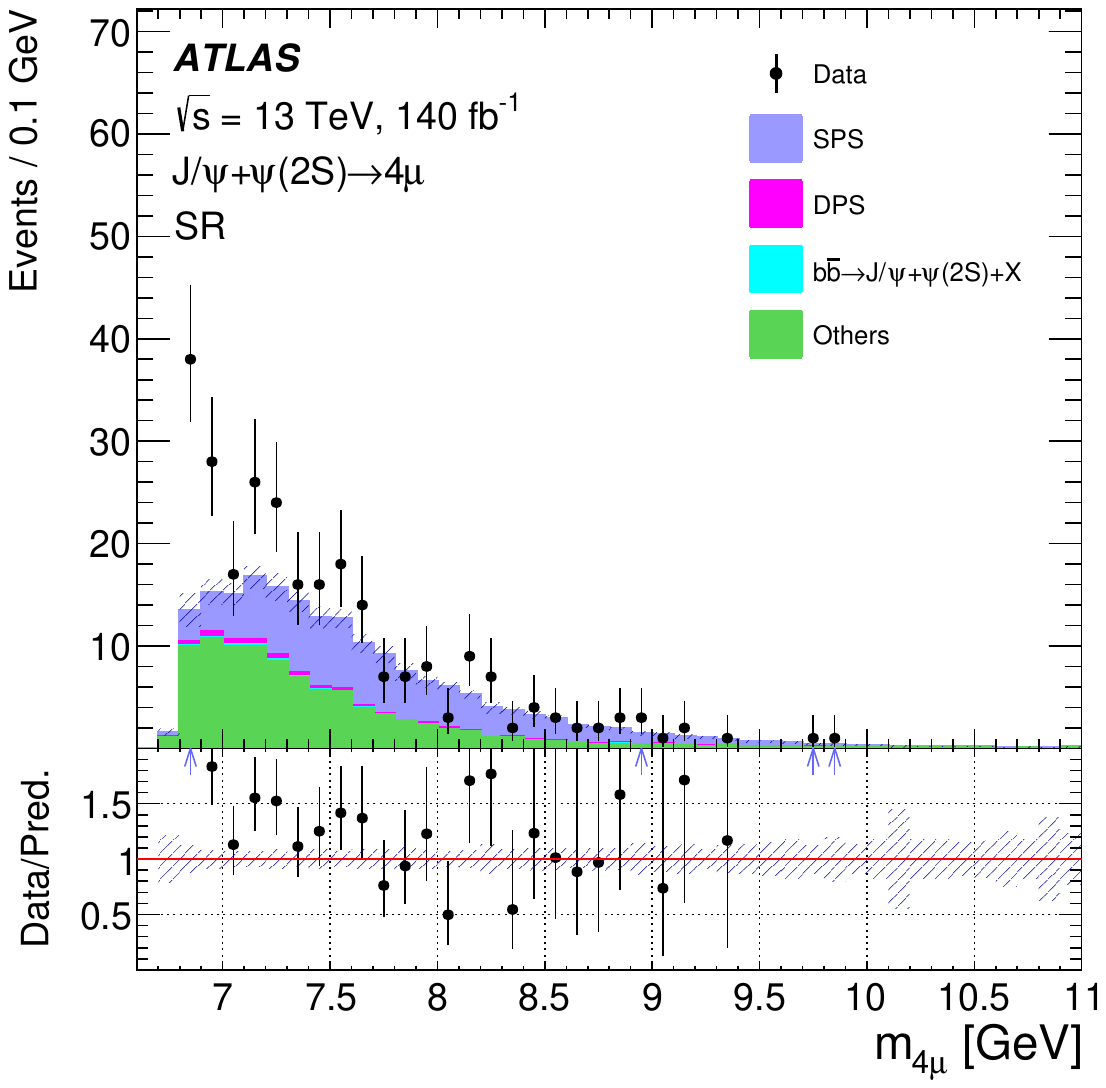}
\includegraphics[width=0.45\textwidth]{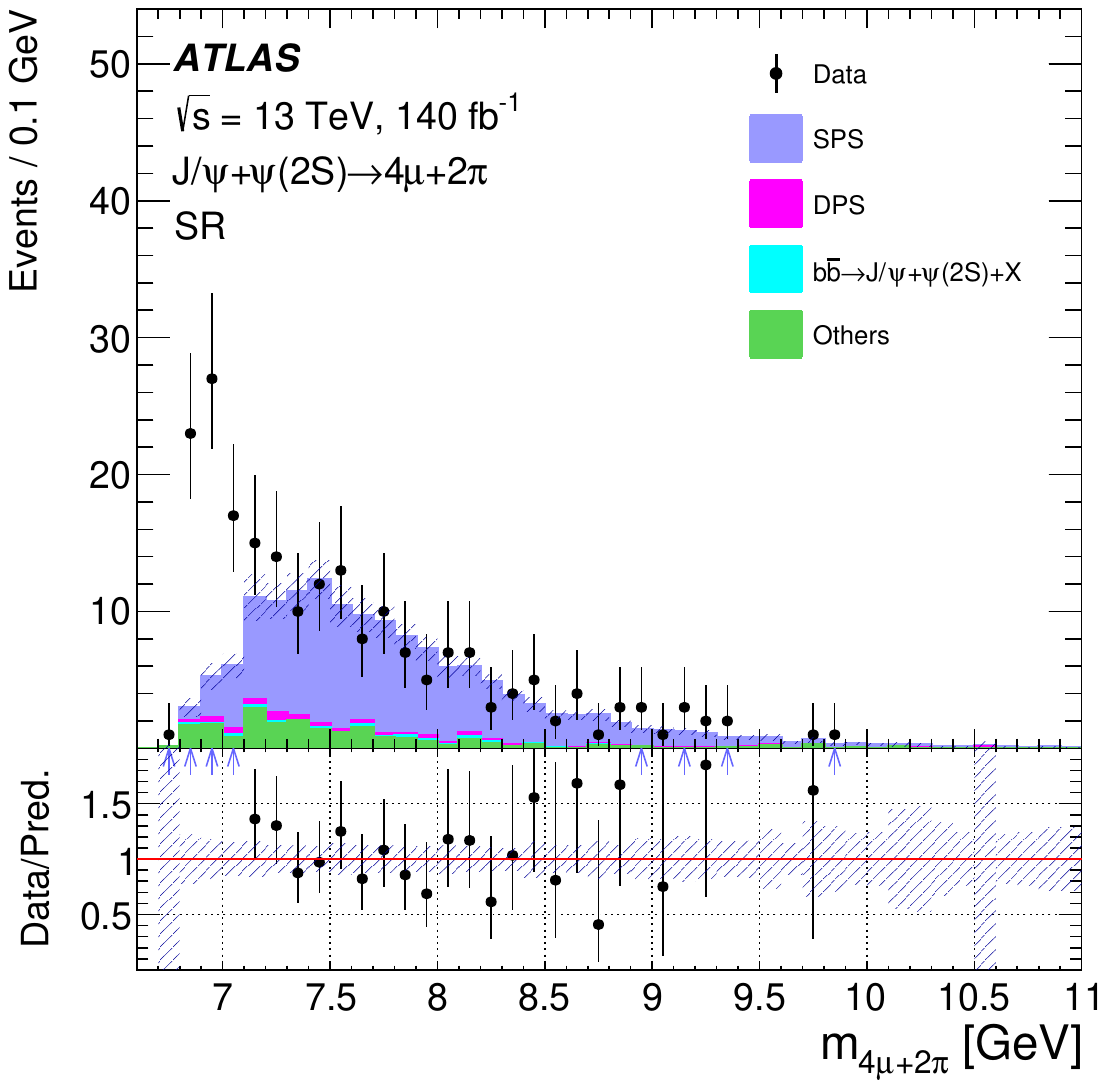}
\caption{The $J/\psi$+$\psi$(2S) mass spectra in the CRs (top) and SRs (bottom) of the $4\mu$ (left) and $4\mu+2\pi$ (right) channels before the fit. The excesses observed in the signal regions arise from the signal contribution. $b\bar{b}\to \text{charmonia}+X$ represents the Nonprompt background. The error bars (shaded areas) represent uncertainties in the data (background predictions) in each bin. Arrows in the lower panels indicate bins in which the ratio of data to the background prediction falls outside the displayed range.}
\label{fig:SRCR}
\end{figure}

The $J/\psi$+$\psi$(2S) mass spectra in the SRs and CRs, as defined in Table~\ref{tab:selection}, with estimated backgrounds prior to the fit, are shown in Figure~\ref{fig:SRCR}. Noticeable excesses of data over the expected background are observed in both SRs. The SPS mass shape is also validated in the CRs, particularly near the di-charmonium production threshold. Since the signal and background cross sections are not measured in this analysis (which would require many theoretical assumptions), their absolute yields and efficiencies are not provided. Instead, the fractions of the four background components in the SRs are listed in Table~\ref{tab:bkgFrac}.

\begin{table*}[htb]
\caption{The fractions of the four background components in the SRs.}
\begin{center}
\begin{tabular}{c|c|c|c|c}
\toprule
Channel & Others & Nonprompt & DPS & SPS \\ \midrule
$4\mu$ & $47.3\%$ & $0.8\%$ & $2.5\%$ & $49.4\%$ \\
$4\mu+2\pi$ & $13.5\%$ & $1.4\%$ & $3.6\%$ & $81.5\%$ \\
\bottomrule
\end{tabular}
\label{tab:bkgFrac}
\end{center}
\end{table*}


\section{Fit model}
\label{sec:fit_model}

Unbinned maximum likelihood fits are performed to extract the signal information from data in the $m_{4\mu}$ and $m_{4\mu+2\pi}$ mass spectra. The likelihood function used in the fit is
\begin{equation}
\label{eq:eq6}
\mathcal{L} = \mathcal{L}_{SR}\left(\bm{\theta},\bm{\lambda}\right) \cdot \mathcal{L}_{CR}\left(\bm{\theta}\right) \cdot \prod_{j=1}^{K} G\left(\theta_j^\prime;\theta_j,\sigma_j\right),
\end{equation}
where $\mathcal{L}_{SR}$ and $\mathcal{L}_{CR}$ are the likelihoods in the SRs and CRs defined in Section~\ref{sec:sel}, $\bm{\lambda}$ are the parameters of interest, $\theta_j$ are nuisance parameters (NP) accounting for systematic uncertainties shared between the SRs and CRs.
Each NP follows a Gaussian distribution having a subsidiary measurement $\theta_j^\prime$, with mean $\theta_j$ and width $\sigma_j$ from pre-fit central values and uncertainties.
Background yields in the $4\mu$ CR, $4\mu+2\pi$ OS CR, and SS CR-like region are used to constrain the background yields in the corresponding $4\mu$ SR, $4\mu+2\pi$ SR, and SS SR-like region, respectively, in a simultaneous fit (a single fit combining the three mass distributions in the SRs and the three background yields in the CRs). Each constraint is implemented as a ratio of total background yields in an SR to that in its corresponding CR, as predicted by pre-fit background estimations.
The signal mass and width are shared in the SRs. %
The feed-up background is assumed to have the same shape and yield in $4\mu+2\pi$ OS SR and SS SR-like. The shape is taken from MC, while the yield is extracted from the fit.
For fits that also involve the di-$J/\psi$ channel, the corresponding likelihood from Ref.~\cite{BPHY-2022-01} is multiplied with Eq.~\ref{eq:eq6} and the statistical inputs from that reference are used. These inputs include the selected data, the background and signal modelling, and the associated nuisance parameters.
Consequently, the di-$J/\psi$ channel contributes an additional fit region when combined with the $J/\psi+\psi\text{(2S)}$ channels. Common parameters, such as those describing the signal or the same systematic source, are simultaneously determined from the fits in all regions.
The list of nuisance parameters implemented for a particular model is given in Appendix~\ref{app:NP_modelC}.

The signal resonances can be either standalone (no interference with the background or other resonances) or interfering, in which case they can interfere with each other or with the SPS background.
Thus, in this paper, three representative signal models are considered, where a relativistic Breit–Wigner (BW) function is used to parameterize the propagator for a resonance. The width in the BW denominator accounts for all decay channels accessible to the resonance, while the numerator is related to the production process.
The three models differ in the number of BWs included and in the way they interfere.
Models $A$~\cite{BPHY-2022-01, CMS-BPH-21-003} and $B$~\cite{LHCb_TQ, BPHY-2022-01} assume that the $X$(6900) state observed in the di-$J/\psi$ channel also decays into $J/\psi$+$\psi$(2S).
In model $A$, the signal probability density functions (PDF) consist of two interfering S-wave BW resonances, which read
\begin{equation}
\begin{aligned}
\label{eq:eq7}
f_{s,\text{di-}J/\psi}^A(x) = \left|\sum_{i=1}^{2}\frac{z_i}{m_i^2 - x^2 -i m_i \Gamma_i(x)}\right|^2 \sqrt{ 1-\frac{4 M_{J/\psi}^2}{x^2} } \otimes R(\bm{\theta}), \\
f_{s,J/\psi+\psi(\text{2S})}^A(x) = \left|\sum_{i=1}^{2}\frac{z_i}{m_i^2 - x^2 -i m_i \Gamma_i(x)}\right|^2 \sqrt{ 1-\left( \frac{M_{J/\psi}+M_{\psi(2S)}}{x} \right)^2 } \otimes R(\bm{\theta}),
\end{aligned}
\end{equation}
where $x$ is the event-by-event reconstructed di-charmonium mass, $m_i$ and $\Gamma_i(x)$ are resonance mass and width
\footnote{For S-waves, $\Gamma_i(x)=\Gamma_i \frac{m_i}{x}\frac{q}{q_i}$, where $q$ ($q_i$) is the momentum of one charmonium in the rest frame of the di-charmonium system at the invariant mass equal to $x$ ($m_i$)~\cite{Zyla:2020zbs}.}
parameters ($m_1 < m_2$), $M$ represents the nominal charmonium mass~\cite{Zyla:2020zbs}, $z_i$ are complex numbers representing the relative magnitudes and phases ($z_1$ is fixed to unity with a zero phase without loss of generality), and $R(\theta)$ is the mass resolution function convolved ($\otimes$) with the BWs. The function is a sum of three Gaussians with different widths and a common mean.
Model $B$ has two BWs -- one interferes with SPS, while the other, $X$(6900), remains standalone. The PDFs in the two channels read
\begin{equation}
\begin{aligned}
\label{eq:eq9}
f_{s+\text{SPS},\text{di-}J/\psi}^B(x) = \left(\left|\frac{z_1}{m_1^2 - x^2 -i m_1 \Gamma_1(x)} + A(x) e^{i\phi_{\text{SPS}}} \right|^2 + \left|\frac{1}{m_2^2 -x^2 -i m_2 \Gamma_2(x)}\right|^2\right) \sqrt{ 1-\frac{4 M_{J/\psi}^2}{x^2} } \otimes R(\bm{\theta}),\\
f_{s,J/\psi+\psi(\text{2S})}^B(x) = \left|\frac{1}{m_2^2 - x^2 -i m_2 \Gamma_2(x)}\right|^2 \sqrt{ 1-\left( \frac{M_{J/\psi}+M_{\psi(2S)}}{x} \right)^2 } \otimes R(\bm{\theta}),
\end{aligned}
\end{equation}
where $A(x)$ and $\phi_{\text{SPS}}$ are the SPS background amplitude and phase relative to the resonance at $m_1$. The term $|A(x)|^2$ reproduces the non-interfering SPS background from the MC prediction. %

The signal yields of $X(6900)$\footnote{For model $A$, the signal yield of $X(6900)$ corresponds to its own squared BW function, without interference with other resonances taken into account.} in the $\text{di-}J/\psi \to 4\mu$ ($N_{\text{di-}J/\psi}$), $J/\psi+\psi(\mathrm{2S}) \to 4\mu$ ($N_{4\mu}$), and $J/\psi+\psi(\mathrm{2S}) \to 4\mu+2\pi$ ($N_{4\mu+2\pi}$) channels are related through the decay branching fractions and the corresponding signal efficiencies. These relationships are expressed by the ratios $f_1$ and $f_2$,

\begin{equation}
\label{eq:eq9b}
\begin{aligned}
f_1 = \frac{N_{4\mu}}{N_{\text{di-}J/\psi}}
= R \,
\frac{\mathcal{B}\!\left(\psi(\mathrm{2S}) \to \mu^+\mu^-\right)\,
\varepsilon_{4\mu}}{\mathcal{B}\!\left(J/\psi \to \mu^+\mu^-\right)\,
\varepsilon_{\text{di-}J/\psi}}, \\[6pt]
f_2 = \frac{N_{4\mu}}{N_{4\mu+2\pi}}
= \frac{\mathcal{B}\!\left(\psi(\mathrm{2S}) \to \mu^+\mu^-\right)\,
\varepsilon_{4\mu}}{\mathcal{B}\!\left(\psi(\mathrm{2S}) \to J/\psi\,\pi^+\pi^-\right)\,
\mathcal{B}\!\left(J/\psi \to \mu^+\mu^-\right)\,
\varepsilon_{4\mu+2\pi}} \, ,
\end{aligned}
\end{equation}

where $\mathcal{B}$ and $\varepsilon$ denote the relevant charmonium decay branching fractions and the corresponding signal efficiencies, respectively.
The parameter
\begin{equation}
\label{eq:eq10}
R \equiv
\frac{\Gamma_{X(6900)\to J/\psi+\psi(\mathrm{2S})}}{\Gamma_{X(6900)\to \text{di-}J/\psi}}
\end{equation}
is the ratio of partial decay widths of $X(6900)$.

Based on MC simulations, the value of $f_1$ for equal partial decay widths ($R=1$) is $0.145 \pm 0.016$, while $f_2 = 0.880 \pm 0.072$. The quoted uncertainties include both statistical and systematic components. In the fit, $R$ is treated as a free parameter.

In model $C$, only $J/\psi$+$\psi$(2S) channels are considered in the simultaneous fit, and the signal PDF consists of a single BW function, corresponding to the second line of Eq.~\ref{eq:eq9}.

To test the potential existence of a second resonance near 7.2~GeV ($X(7200)$) in the $J/\psi+\psi(\text{2S})$ channels, a standalone $X(7200)$ component is added in each of the models described above. A separate fit is then performed for each extended model.
The mass and width of this resonance are fixed at $7.22 $~\GeV and $0.09 $~\GeV, respectively, as determined in the previous ATLAS publication~\cite{BPHY-2022-01}. %
The ratio of signal yields of $X$(7200) and $X$(6900), $r = N_{X(7200)}/N_{X(6900)}$, is treated as a free parameter in the fit.
%


\section{Results}
\label{sec:res}

\begin{figure}[htbp]
\centering
\includegraphics[width=0.45\textwidth]{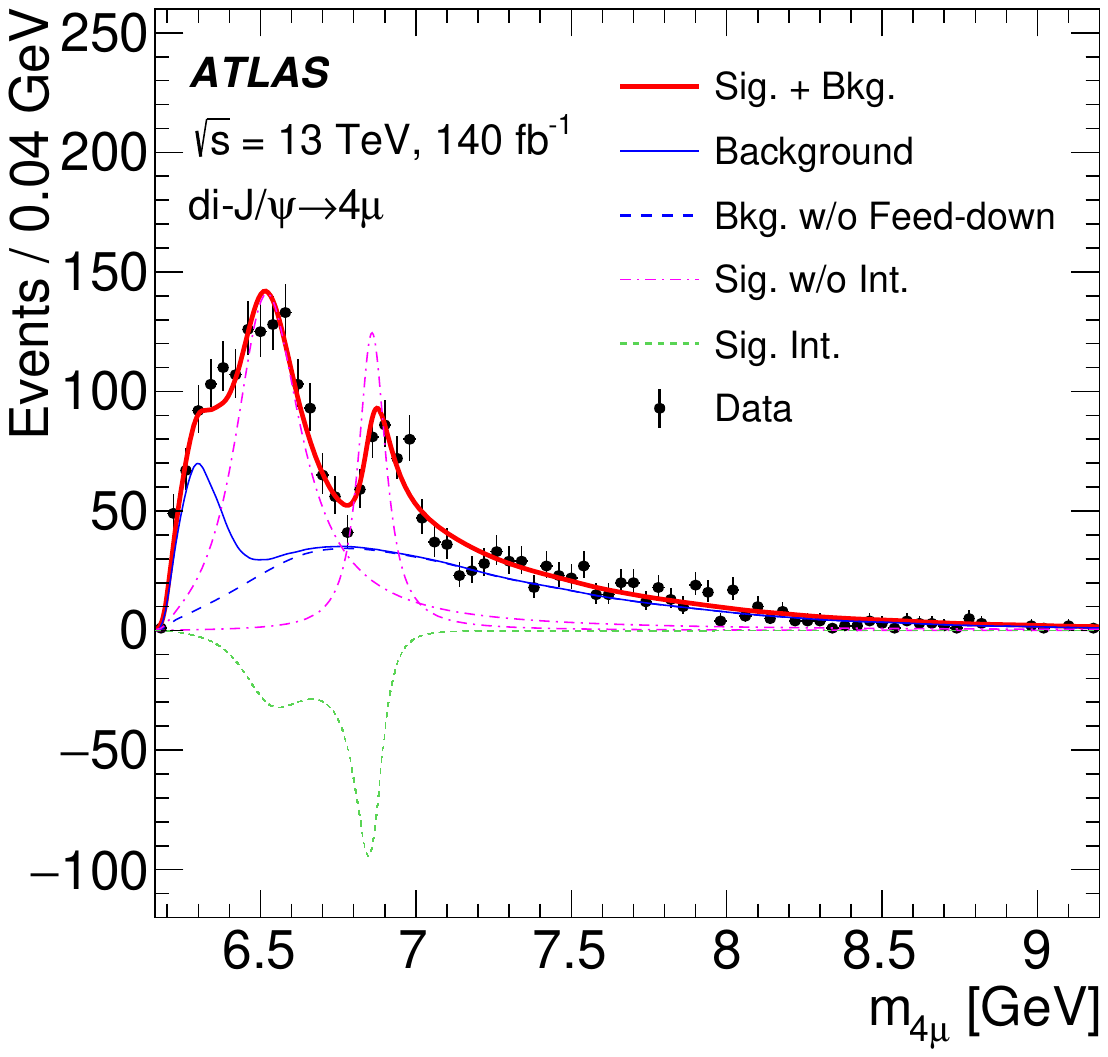}
\includegraphics[width=0.45\textwidth]{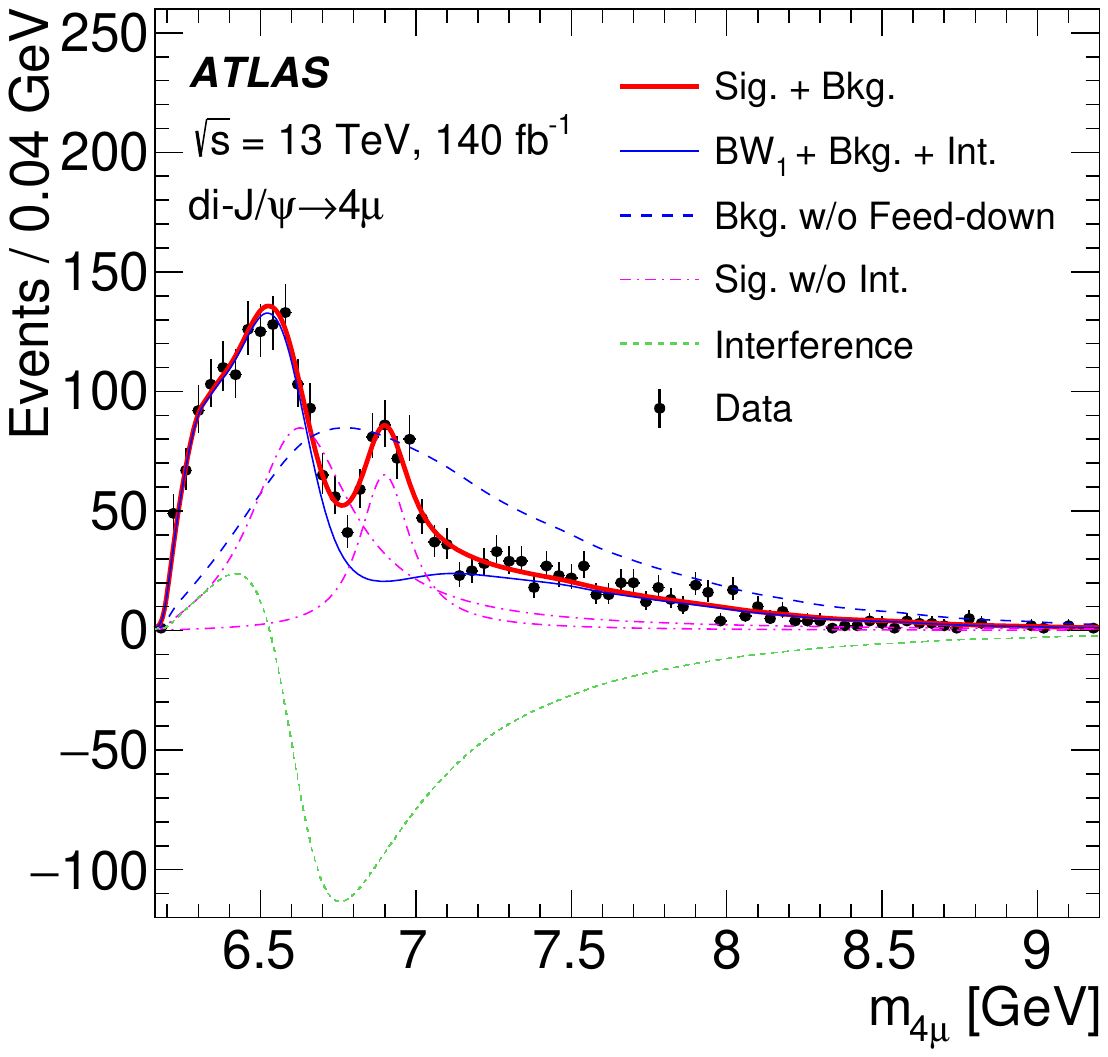}
\includegraphics[width=0.45\textwidth]{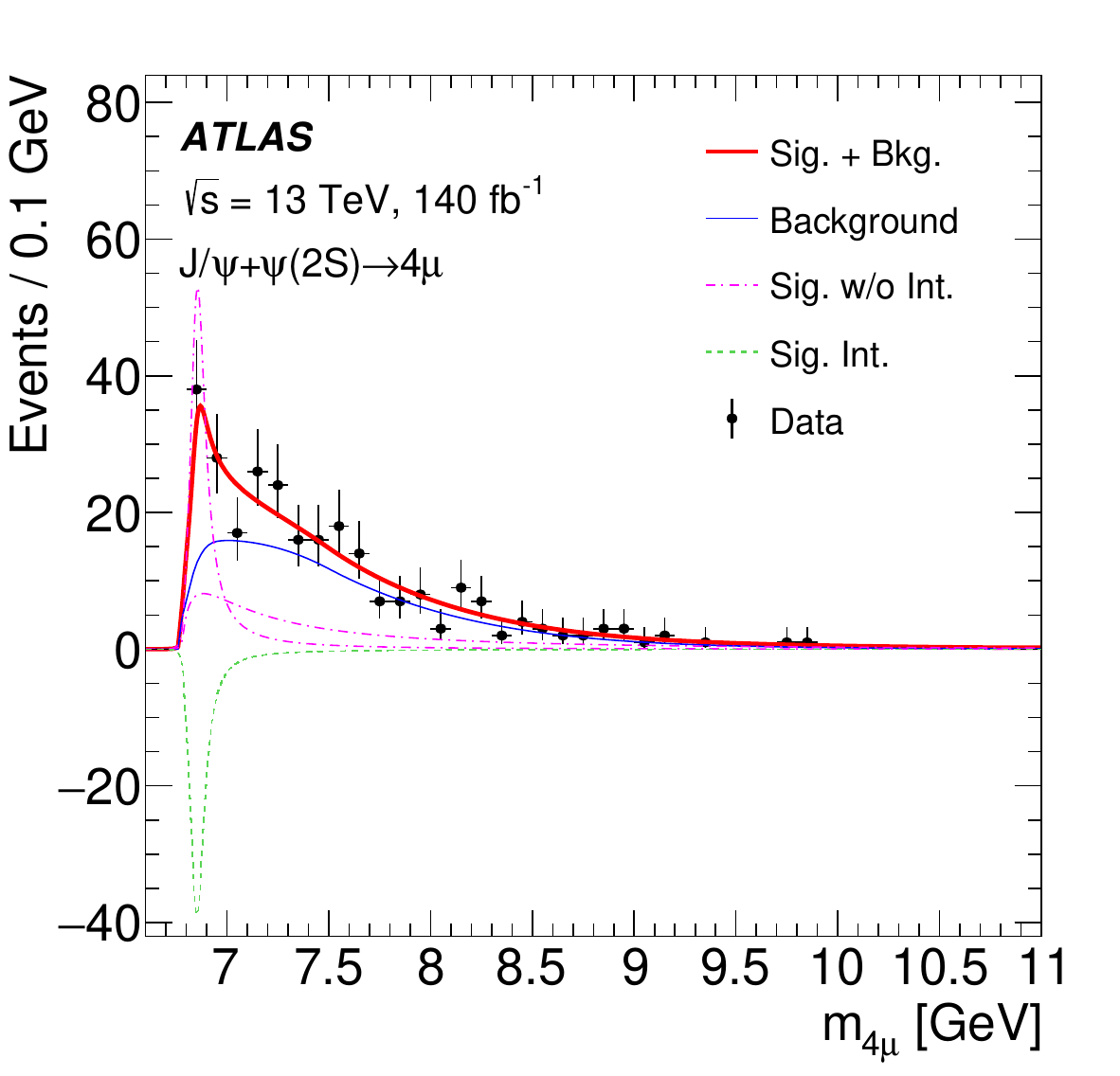}
\includegraphics[width=0.45\textwidth]{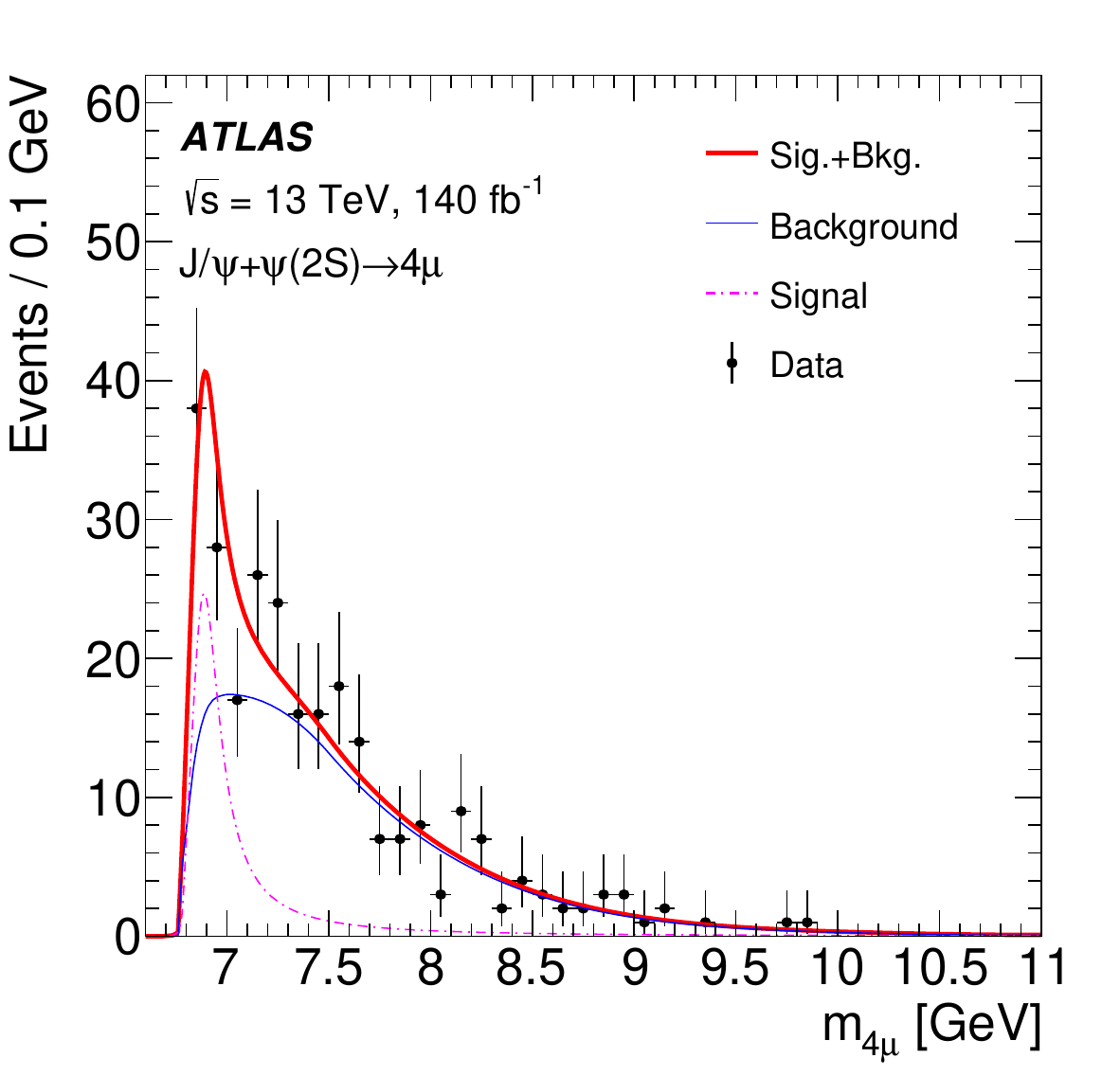}
\includegraphics[width=0.45\textwidth]{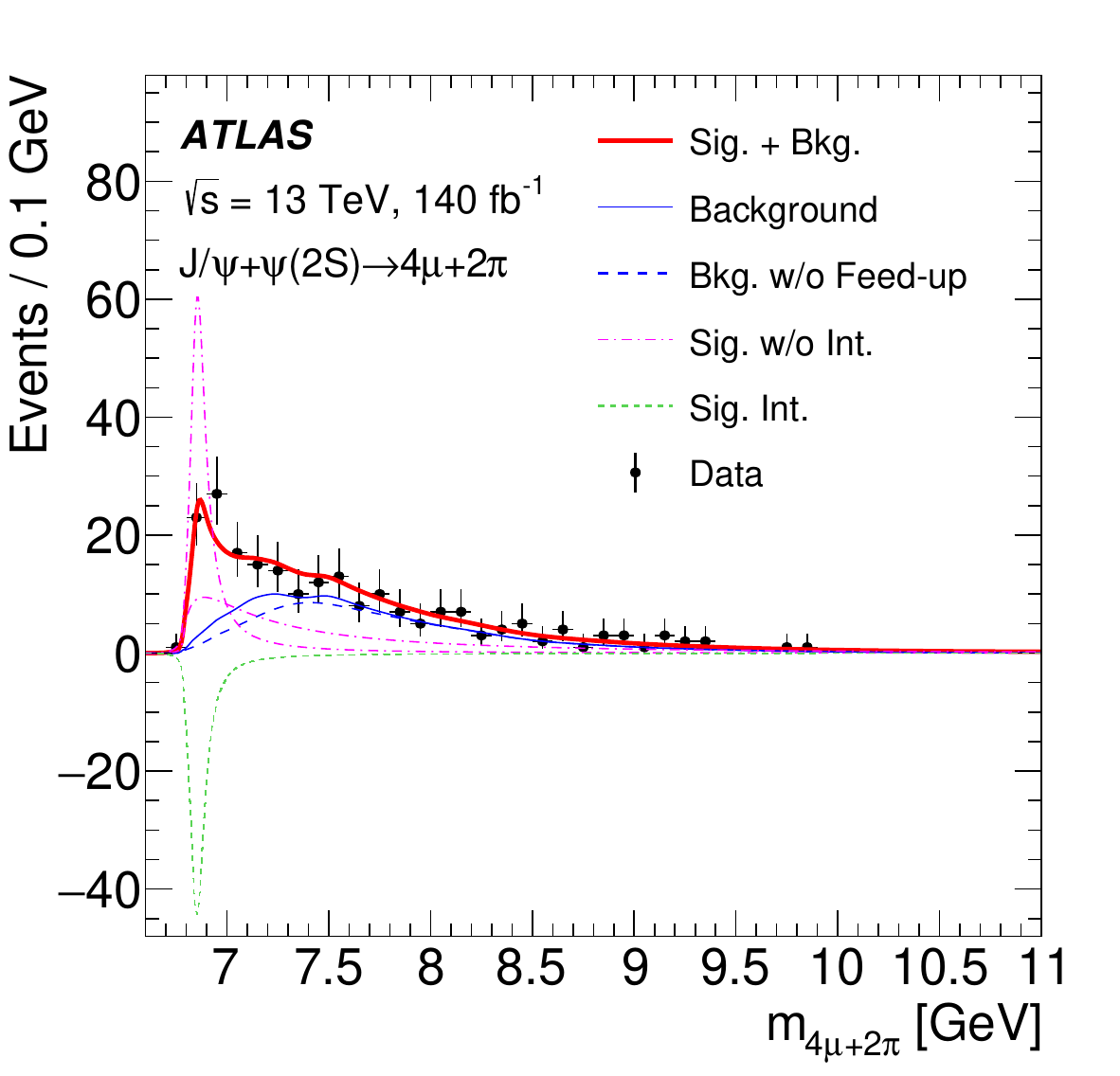}
\includegraphics[width=0.45\textwidth]{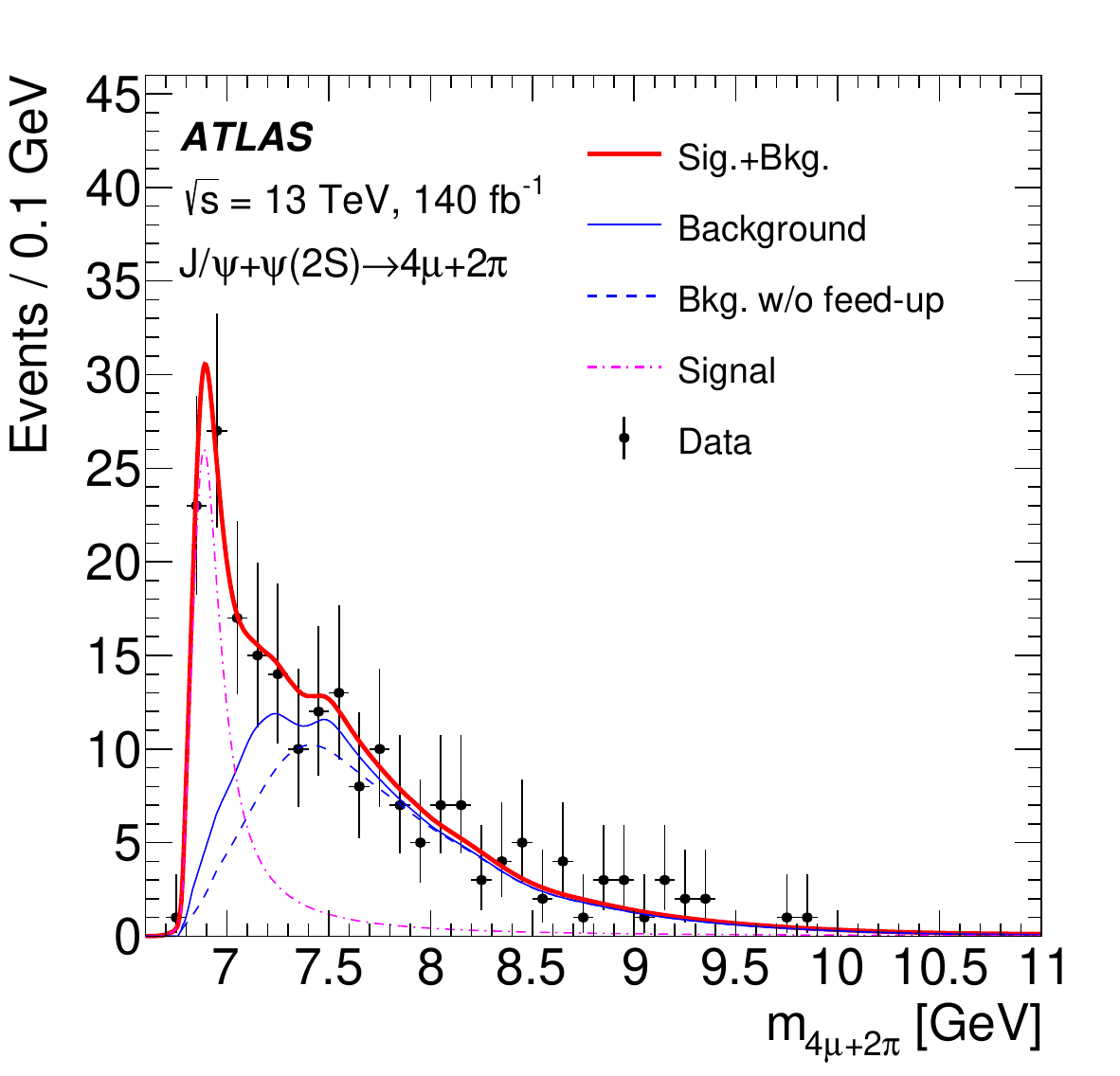}
\caption{Fits to the di-$J/\psi$ and $J/\psi+\psi$(2S) mass spectra in the SRs for model $A$ (left) and $B$ (right). For each model, the di-$J/\psi$ spectrum (top) and the $J/\psi+\psi$(2S) spectra in the $4\mu$ (middle) and $4\mu+2\pi$ (bottom) channels are shown. The purple dash-dotted lines indicate the components of individual resonances, while the green short dashed ones represent the interferences. Note that no interference is present in the $J/\psi+\psi$(2S) channels for model $B$ since $X$(6900) is standalone.
}
\label{fig:modelAandB}
\end{figure}

\begin{figure}[htbp]
\centering
\includegraphics[width=0.45\textwidth]{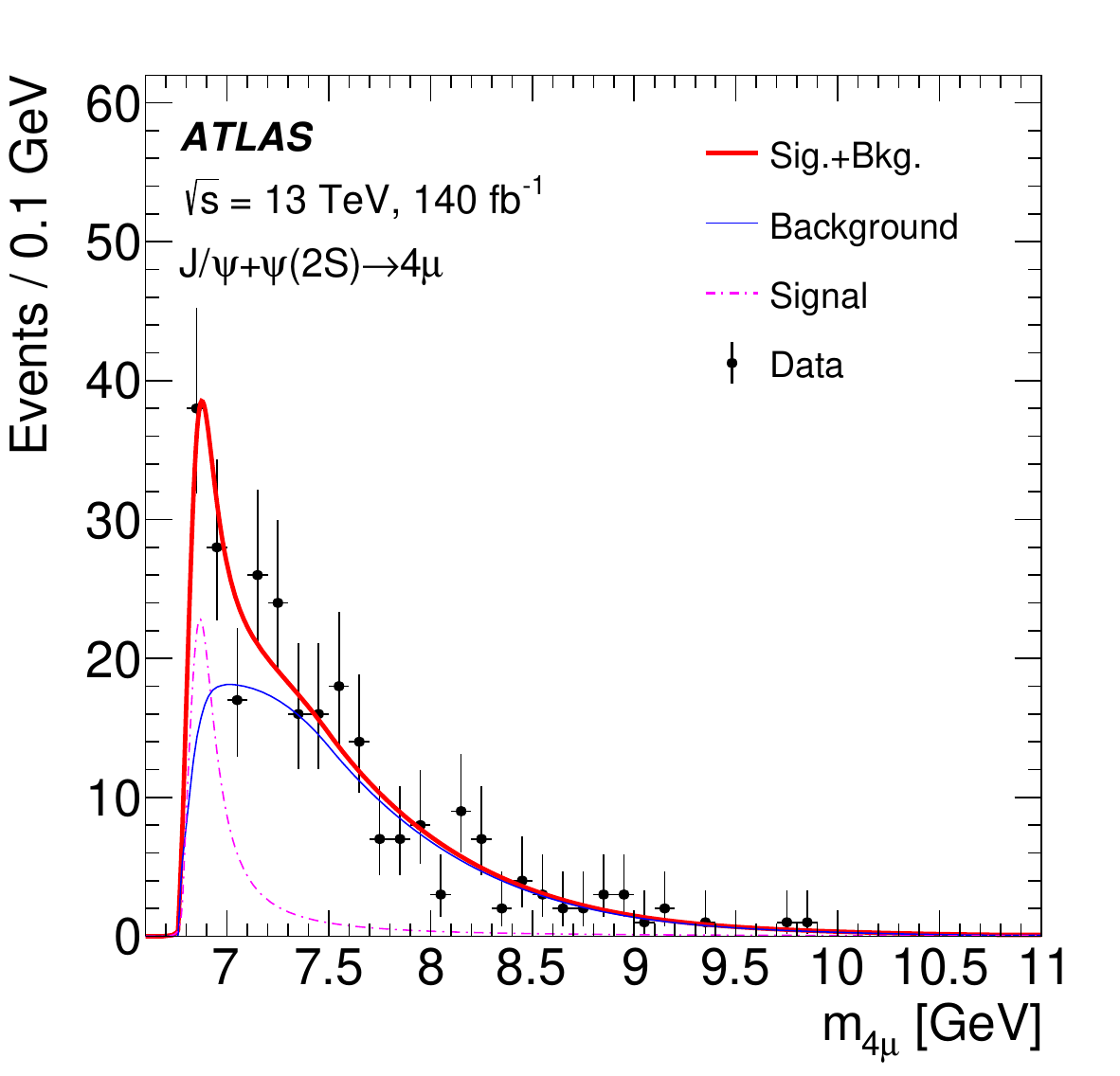}\\
\includegraphics[width=0.45\textwidth]{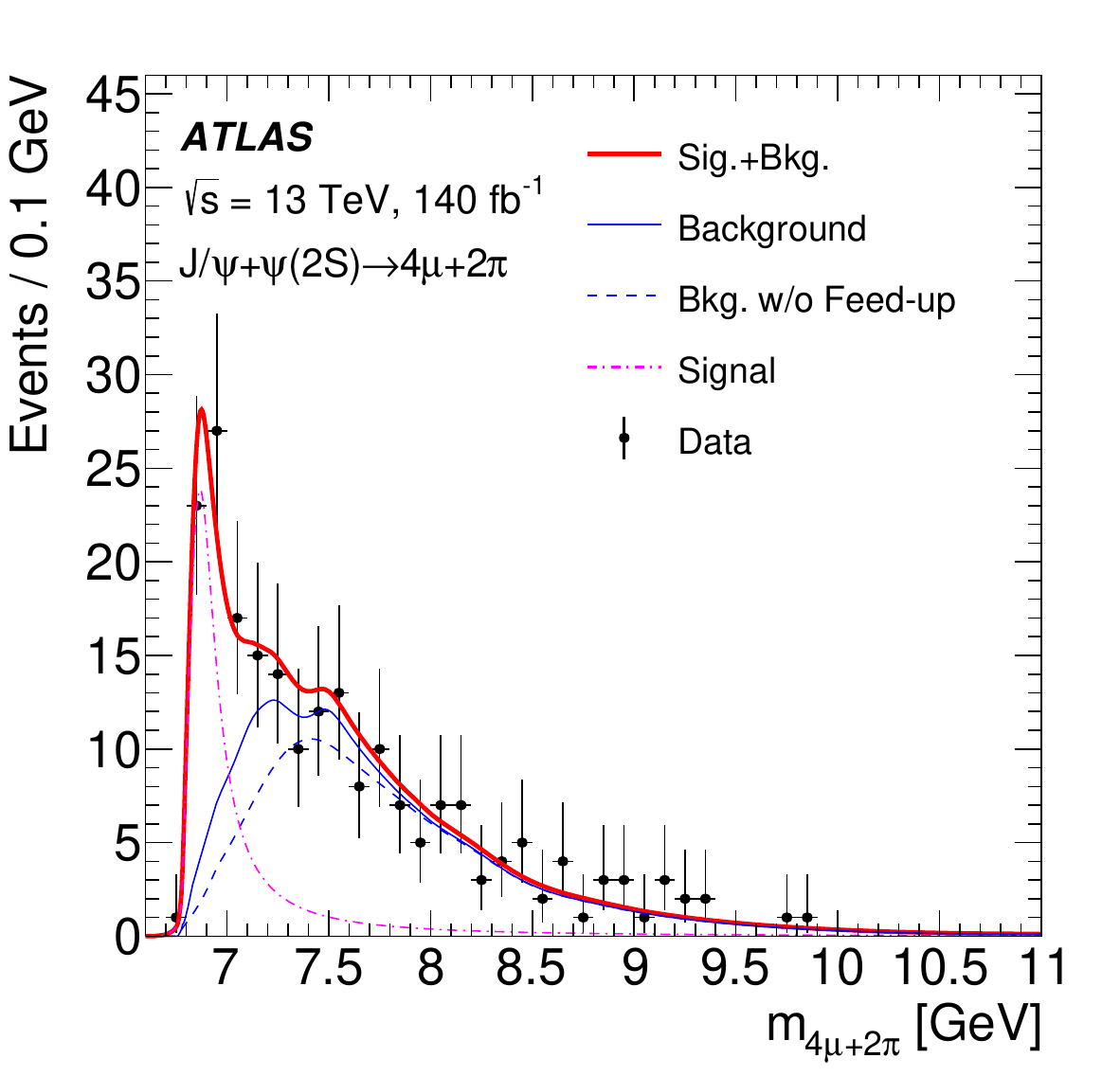}
\includegraphics[width=0.45\textwidth]{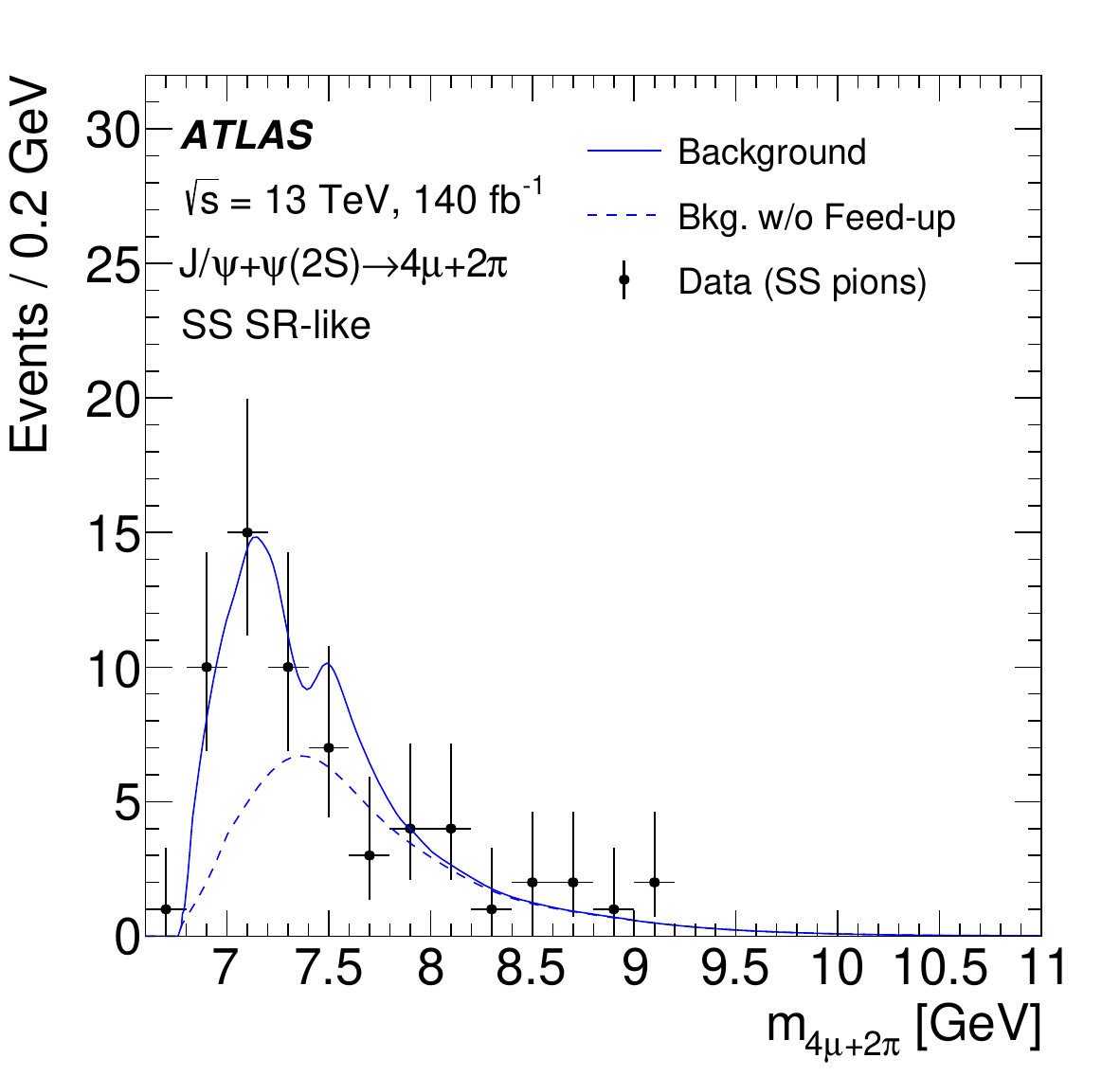}
\caption{Fits to the $J/\psi$+$\psi$(2S) mass spectra with model $C$ in the $4\mu$ SR (top), $4\mu+2\pi$ SR (bottom left), and $4\mu+2\pi$ SS SR-like region (bottom right). The purple dash-dotted lines represent the signal resonances.}
\label{fig:unb_mass_fit}
\end{figure}

The fitted di-$J/\psi$ and $J/\psi$+$\psi$(2S) mass spectra for models $A$ and $B$ are shown in Figure~\ref{fig:modelAandB}.
The fitted $J/\psi$+$\psi$(2S) mass spectra using model $C$ are shown in Figure~\ref{fig:unb_mass_fit}, which also shows the fitted spectrum in the SS SR-like region with $\Delta R < 0.25$. The fit results in this region obtained with the other two models are similar, and are therefore not displayed in Figure~\ref{fig:modelAandB}.
The fitted masses and widths of $X$(6900), and the ratio of partial widths $R$ are given in Table \ref{tab:mass_width1}.
A discussion of systematics uncertainties is presented in Section~\ref{sec:sys}.
Taking model $A$ as the nominal scenario and model $B$ to assess systematic effects, the ratio of partial decay widths is found to be $R = 1.08\pm 0.20_{-0.17}^{+0.40}$. The fitted resonance mass is consistently around 6.9~\GeV across all three models. The fitted width is about 0.08~\GeV with model $A$, and about 0.18~\GeV with model $B$ and $C$.
The fitted parameters of the near-threshold resonance in the di-$J/\psi$ channel are listed in Table~\ref{tab:mass_width2}.
The signal significance of X(6900) for model $C$ with signal shape parameters fixed to their best-fit values reaches 8.9$\sigma$
\footnote{The asymptotic formula based on the profile likelihood ratio, $Z = \sqrt{2\ln \frac{\mathcal{L}(\hat{s},\hat{\bm{\theta}})}{\mathcal{L}(0,\hat{\hat{\bm{\theta}}})}}$, is used to calculate the overall significance, where $s$ is the signal yield and $\theta$ are NPs~\cite{Cowan:2010js}. In the calculations, all signal shape parameters are fixed to their best-fit values.}.
Compared to Ref.~\cite{BPHY-2022-01}, more precise mass and width are determined in the $J/\psi$+$\psi$(2S) channels. The results from model $C$ are consistent with those obtained from the individual fits to the $4\mu$ and $4\mu+2\pi$ channels.
By including $X$(7200) in model $C$, $r$ is found to be $0.12\pm 0.11$, with an upper limit of $0.41$ at 95\% confidence level using the $CL_s$ method~\cite{Read:2002hq} and accounting for variations in the mass and width of $X$(7200) from Ref.~\cite{BPHY-2022-01}. The inclusion of $X$(7200) in models A and B yields consistent results.
An alternative fit using model $C$, in which the SRs and CRs are merged, is also performed. The fitted resonance mass and width obtained from this approach are consistent with those from the nominal fit (Appendix~\ref{app:alt_fit}), with a slightly decreased significance as expected due to the larger background component.

\begin{table}[htbp]
\caption{ The fitted masses and natural widths of $X$(6900) obtained with three fit models are listed. The ratio of partial widths of Eq.~\ref{eq:eq10} is also given for model $A$ and $B$. The first uncertainty is statistical and the second is systematic. }
\label{tab:mass_width1}
\centering
\renewcommand{\arraystretch}{1.2}
\begin{tabular}{cccc}
\toprule
& model $A$ & model $B$  & model $C$  \\ \midrule
$m_2$ / \GeV      & $6.860\pm 0.023\pm 0.010$ & $6.902\pm 0.008\pm 0.010$ & $6.884\pm 0.017_{-0.005}^{+0.058}$ \\
$\Gamma_2$ / \GeV & $0.082\pm 0.032\pm 0.015$ & $0.183\pm 0.025\pm 0.007$ & $0.178\pm 0.054_{-0.024}^{+0.176}$ \\
$R$      & $1.08\pm 0.20_{-0.09}^{+0.40}$ & $0.93\pm 0.17\pm 0.11$ & --- \\
\bottomrule
\end{tabular}
\end{table}

\begin{table}[htbp]
\caption{ The fitted masses and natural widths of the near-threshold resonance in the di-$J/\psi$ channel for models $A$ and $B$. The first uncertainty is statistical and the second is systematic. }
\label{tab:mass_width2}
\centering
\renewcommand{\arraystretch}{1.2}
\begin{tabular}{ccc}
\toprule
& model $A$ & model $B$ \\ \midrule
$m_1$ / \GeV      & $6.531\pm 0.011\pm 0.002$ & $6.646\pm 0.018\pm 0.021$ \\
$\Gamma_1$ / \GeV & $0.244\pm 0.021\pm 0.021$ & $0.390\pm 0.045\pm 0.037$ \\
\bottomrule
\end{tabular}
\end{table}


\section{Systematic uncertainties}
\label{sec:sys}

Systematic uncertainties arise from signal modeling, background estimation and detector effects.
Signal modelling uncertainties include variations of the signal resonance's natural decay width, the effect of turning on or off the correction on its $p_{\text{T}}$, as well as uncertainties in the charmonium decay branching fractions~\cite{Zyla:2020zbs}.
Detector-related uncertainties include the muon momentum calibration, trigger, reconstruction and identification efficiencies. A systematic uncertainty is assigned due to the use of mass-independent fixed-width resolution functions.
Background uncertainties account for the impact of statistical fluctuations in the background MC and the correction of the di-charmonium $p_{\text{T}}$ distribution on its mass shape parameters for the total backgrounds in the SRs.
The dominant shape systematic uncertainties affecting the SPS background arise from limited MC statistics and the modeling uncertainty associated with the \Pythia parameter \texttt{pT0timesMPI}, as described in Section~\ref{sec:common_backgrounds}. To evaluate the impact of the latter, the \texttt{pT0timesMPI} parameter is varied to produce alternative MC templates, which are transformed into a shape systematic uncertainty for the background. The nominal value and associated uncertainty of this parameter is $0.61\pm 0.05$.
In the $4\mu$ channel, a mass shape non-closure uncertainty is included for the Others background. In the $4\mu+2\pi$ channel, the mass shape uncertainty of the feed-up background is evaluated by varying the fitted signal parameters obtained from the di-$J/\psi$ channel.
Additional systematic uncertainties arise from the inclusion of $X$(7200) and potential biases from the fit, as determined from toy MC studies.
The S-wave BW functions are replaced by P-wave or D-wave~\cite{BlattWeisskopf:1952,VonHippel:1972fg} to test the variations of fitted parameters, which contributes the largest uncertainty for $X$(6900)'s width in model $C$.
When combining with the di-$J/\psi$ channel, the same systematic uncertainties are correlated with the $J/\psi$+$\psi$(2S) channels.
The breakdown of uncertainties from different sources for the parameters of $X$(6900) in all three models is provided in Table~\ref{tab:sys_mass_width}.

\begin{table}[htbp]
\caption{ Different sources of uncertainties in the mass and natural width (in \MeV) of $X$(6900), and the ratio of partial decay widths of Eq.~\ref{eq:eq10} for the three fit models. Entries with long dashes mean the uncertainty is negligible. }
\label{tab:sys_mass_width}
\centering
\renewcommand{\arraystretch}{1.2}
\begin{tabular}{c|ccc|ccc|cc}
\toprule
& \multicolumn{3}{c|}{Model $A$} & \multicolumn{3}{c|}{Model $B$} & \multicolumn{2}{c}{Model $C$} \\ \cmidrule{2-9}
Sources & $m$ & $\Gamma$ & $R$ & $m$ & $\Gamma$ & $R$ & $m$ & $\Gamma$ \\ \midrule
Statistical uncertainties  & $\pm 23$  & $\pm 32$  & $\pm 0.20$ & $\pm 8.1$ & $\pm 25$ & $\pm 0.17$ & $\pm 17$ & $\pm 54$ \\
Di-$J/\psi$ bkg. modeling  & $\pm 5.4$  & $\pm 7.7$ & $\pm 0.041$ & $\pm 1.4$ & ---  & --- & ---  & --- \\
$J/\psi$+$\psi$(2S) bkg. modeling & $\pm 4.1$ & $\pm 6.7$ & $\pm 0.038$ & --- & --- & --- & $\pm 2.6$  & $\pm 8.6$ \\
Signal ratios              & ---  & ---  & $\pm 0.069$ & --- & --- & $\pm 0.054$ & --- & --- \\
Mass resolution            & $\pm 6.8$  & $\pm 8.6$ & $\pm 0.033$ & $\pm 1.3$ & $\pm 2.2$ & $\pm 0.002$ & $\pm 0.3$  & $\pm 4.4$ \\
Feed-up bkg.               & --- & --- & $\pm 0.002$ & --- & --- & $\pm 0.002$ & $\pm 0.3$  & $\pm 0.6$ \\
P or D-wave BW             & $\pm 3.2$  & $\pm 5.4$ & $+0.39$ & $\pm 9.9$ & $\pm 2.9$  & $\pm 0.094$ & $+58$ & $+170$ \\
Fit bias                   & $\pm 0.2$  & $\pm 2.1$ & $\pm 0.007$ & $\pm 0.2$ & $\pm 4.7$  & $\pm 0.010$ & $\pm 0.4$  & $\pm 11$ \\
Inclusion of $X$(7200)     & $\pm 1.0$  & $\pm 1.2$ & --- & $\pm 0.1$ & $\pm 3.8$  & $\pm 0.001$ & $\pm 4.2$  & $\pm 19$ \\
\bottomrule
\end{tabular}
\end{table}


%

\section{Summary}
\label{sec:sum}

To better understand the excess observed in the $J/\psi$+$\psi$(2S) mass spectrum~\cite{BPHY-2022-01}, both the $\psi$(2S)$\to\mu\mu$ and $\psi$(2S)$\to J/\psi + 2\pi$ decay modes, the latter studied for the first time in this context, are analyzed using $pp$ collision data collected by the ATLAS experiment at $\sqrt{s} = 13$~\TeV, corresponding to an integrated luminosity of 140 fb$^{-1}$.
The fitted resonance mass is consistent across all three models.
The mass and width of $X$(6900) obtained in model $C$ are consistent between the two $\psi$(2S) decay channels, and also agree with, while superseding, the corresponding values of model $\beta$ in Ref.~\cite{BPHY-2022-01}.
The signal significance reaches 8.9$\sigma$ for the resonance near 6.9~\GeV in the $J/\psi$+$\psi$(2S) channels (compared with 4.7$\sigma$ reported in Ref.~\cite{BPHY-2022-01}).
This result confirms the presence of structures in the $J/\psi$+$\psi$(2S) channels, favoring a single-resonance hypothesis near 6.9~\GeV, while the existence of a potential resonance near 7.2~\GeV is not supported by the current data. All three models provide a reasonable description of the data, with two of them indicating a ratio of partial decay widths for $X$(6900)$\to J/\psi$+$\psi$(2S) and $X$(6900)$\to 2 J/\psi$ near 1. Future data will allow a more precise determination of the nature of this excess, including its underlying model and quantum numbers, as well as the exploration of possible additional resonances.
%


%
%
%
%
%
%
%

%
\clearpage
\section*{Acknowledgements}
%

%
%

%
%

We thank CERN for the very successful operation of the LHC and its injectors, as well as the support staff at
CERN and at our institutions worldwide without whom ATLAS could not be operated efficiently.

The crucial computing support from all WLCG partners is acknowledged gratefully, in particular from CERN, the ATLAS Tier-1 facilities at TRIUMF/SFU (Canada), NDGF (Denmark, Norway, Sweden), CC-IN2P3 (France), KIT/GridKA (Germany), INFN-CNAF (Italy), NL-T1 (Netherlands), PIC (Spain), RAL (UK) and BNL (USA), the Tier-2 facilities worldwide and large non-WLCG resource providers. Major contributors of computing resources are listed in Ref.~\cite{ATL-SOFT-PUB-2026-001}.

We gratefully acknowledge the support of ANPCyT, Argentina; YerPhI, Armenia; ARC, Australia; BMWFW and FWF, Austria; ANAS, Azerbaijan; CNPq and FAPESP, Brazil; NSERC, NRC and CFI, Canada; CERN; ANID, Chile; CAS, MOST and NSFC, China; Minciencias, Colombia; MEYS CR, Czech Republic; DNRF and DNSRC, Denmark; IN2P3-CNRS and CEA-DRF/IRFU, France; SRNSFG, Georgia; BMFTR, HGF and MPG, Germany; GSRI, Greece; RGC and Hong Kong SAR, China; ICHEP and Academy of Sciences and Humanities, Israel; INFN, Italy; MEXT and JSPS, Japan; CNRST, Morocco; NWO, Netherlands; RCN, Norway; MNiSW, Poland; FCT, Portugal; MNE/IFA, Romania; MSTDI, Serbia; MSSR, Slovakia; ARIS and MVZI, Slovenia; DSI/NRF, South Africa; MICIU/AEI, Spain; SRC and Wallenberg Foundation, Sweden; SERI, SNSF and Cantons of Bern and Geneva, Switzerland; NSTC, Taipei; TENMAK, T\"urkiye; STFC/UKRI, United Kingdom; DOE and NSF, United States of America.

Individual groups and members have received support from BCKDF, CANARIE, CRC and DRAC, Canada; CERN-CZ, FORTE and PRIMUS, Czech Republic; COST, ERC, ERDF, Horizon 2020, ICSC-NextGenerationEU and Marie Sk{\l}odowska-Curie Actions, European Union; Investissements d'Avenir Labex, Investissements d'Avenir Idex and ANR, France; DFG and AvH Foundation, Germany; Herakleitos, Thales and Aristeia programmes co-financed by EU-ESF and the Greek NSRF, Greece; BSF-NSF and MINERVA, Israel; NCN and NAWA, Poland; La Caixa Banking Foundation, CERCA and AGAUR programs from Generalitat de Catalunya and PROMETEO and GenT Programmes Generalitat Valenciana, Spain; G\"{o}ran Gustafssons Stiftelse, Sweden; The Royal Society and Leverhulme Trust, United Kingdom; Eric and Wendy Schmidt Fund for Strategic Innovation, United States of America.

In addition, individual members wish to acknowledge support from CERN: European Organization for Nuclear Research (CERN DOCT); Chile: Agencia Nacional de Investigaci\'on y Desarrollo (ANID FONDECYT reg. 1230987, FONDECYT 1230812, FONDECYT 1240864, Fondecyt 3240661, Fondecyt Regular 1240721); China: Chinese Ministry of Science and Technology (MOST-2023YFA1605700, MOST-2023YFA1609300), National Natural Science Foundation of China (NSFC - 12175119, NSFC 12275265); Czech Republic: Czech Science Foundation (GACR - 24-11373S), Ministry of Education Youth and Sports (ERC-CZ-LL2327, FORTE CZ.02.01.01/00/22\_008/0004632), PRIMUS Research Programme (PRIMUS/21/SCI/017); EU: H2020 European Research Council (ERC - 101002463); European Union: European Research Council (BARD No. 101116429, ERC - 948254, ERC 101089007), European Regional Development Fund (HE COFUND GA No.101081355, ERDF), European Union, Future Artificial Intelligence Research (FAIR-NextGenerationEU PE00000013), Italian Center for High Performance Computing, Big Data and Quantum Computing (ICSC, NextGenerationEU), Marie Sklodowska-Curie Actions (ERC Horizon-2020, GA 956086, GAP-101168829); France: Agence Nationale de la Recherche (ANR-21-CE31-0013, ANR-21-CE31-0022, ANR-22-EDIR-0002, ANR-24-CE31-0504-01); Germany: Baden-Württemberg Stiftung (BW Stiftung-Postdoc Eliteprogramme), Deutsche Forschungsgemeinschaft (DFG - 469666862, DFG - CR 312/5-2); China: Research Grants Council (GRF); Italy: Istituto Nazionale di Fisica Nucleare (ICSC, NextGenerationEU), Ministero dell'Università e della Ricerca (NextGenEU 153D23001490006 M4C2.1.1, NextGenEU I53D23000820006 M4C2.1.1, NextGenEU I53D23001490006 M4C2.1.1, SOE2024\_0000023); Japan: Japan Society for the Promotion of Science (JSPS KAKENHI  JP25H0063, JSPS KAKENHI JP22H01227, JSPS KAKENHI JP22H04944, JSPS KAKENHI JP22KK0227, JSPS KAKENHI JP24K23939, JSPS KAKENHI JP24KK0251, JSPS KAKENHI JP25H00650, JSPS KAKENHI JP25H01291, JSPS KAKENHI JP25K01011, JSPS KAKENHI JP25K01023); Norway: Research Council of Norway (RCN-314472); Poland: Ministry of Science and Higher Education (IDUB AGH, POB8, D4 no 9722), Polish National Science Centre (NCN 2021/42/E/ST2/00350, NCN OPUS 2023/51/B/ST2/02507, NCN OPUS nr 2022/47/B/ST2/03059, NCN UMO-2019/34/E/ST2/00393, UMO-2022/47/O/ST2/00148, UMO-2023/49/B/ST2/04085, UMO-2023/51/B/ST2/00920, UMO-2024/53/N/ST2/00869); Portugal: Foundation for Science and Technology (FCT); Spain: Agència de Gestió d'Ajuts Universitaris i de Recerca. (AGAUR - 2023 BP 00141), Generalitat Valenciana (ASFAE/2022/008), Ministry of Science and Innovation (MCIN \& NextGenEU PCI2022-135018-2, MICIN \& FEDER PID2021-125273NB, RYC2019-028510-I, RYC2020-030254-I, RYC2021-031273-I, RYC2022-038164-I), Ministerio de Ciencia, Innovación y Universidades/Agencia Estatal de Investigaci\'on (EU NextGenerationEU (PRTR-C17.I1), PID2022-142604OB-C22); Sweden: Carl Trygger Foundation (Carl Trygger Foundation CTS 22:2312), Swedish Research Council (Swedish Research Council 2023-04654, VR 2021-03651, VR 2022-03845, VR 2022-04683, VR 2023-03403, VR 2024-05451), Knut and Alice Wallenberg Foundation (KAW 2018.0458, KAW 2022.0358, KAW 2023.0366); Switzerland: Swiss National Science Foundation (SNSF - PCEFP2\_194658); United Kingdom: Royal Society (NIF-R1-231091); United States of America: U.S. Department of Energy (ECA DE-AC02-76SF00515), Neubauer Family Foundation.

%
%


%
\clearpage
\appendix
\part*{Appendix}
\addcontentsline{toc}{part}{Appendix}


\section{Nuisance parameters with model $C$}
\label{app:NP_modelC}

The nuisance parameters and their pre-fit values are given in Table~\ref{tab:NP_modelC} for model $C$. The sources of the associated uncertainties are described in Section~\ref{sec:sys}. Parameters related to the shapes of distributions do not have numerical values and are indicated as "shape parameter" in Table~\ref{tab:NP_modelC}.

\begin{table}[htbp]
\caption{ The nuisance parameters and their pre-fit values for model $C$. The ones related to shapes of distributions are listed as shape parameters that do not have quantitative values. }
\label{tab:NP_modelC}
\centering
\renewcommand{\arraystretch}{1.2}
\begin{tabular}{c|c}
\toprule
$4\mu+2\pi$ SS SR-like background shape & shape parameter \\
$4\mu+2\pi$ feed-up background shape & shape parameter \\
Ratio of background in $4\mu+2\pi$ SS CR-like and SR-like & $1.469^{+0.031}_{-0.078}$ \\
$4\mu+2\pi$ OS SR background shape & shape parameter \\
$4\mu+2\pi$ OS SR SPS Pythia parameter & shape parameter \\
Ratio of background in $4\mu+2\pi$ OS CR and SR & $1.295^{+0.028}_{-0.066}$ \\
$4\mu$ SR background shape & shape parameter \\
$4\mu$ SR Others non-closure & shape parameter \\
Ratio of background in $4\mu$ CR and SR & $1.604^{+0.055}_{-0.066}$ \\
Signal mass resolution (muon calibration) & shape parameter \\
Signal mass resolution (mass dependence) & shape parameter \\
Ratio of signal yields in $4\mu$ and $4\mu+2\pi$ & $0.880\pm 0.072$ \\
\bottomrule
\end{tabular}
\end{table}

\section{Alternative fit with model $C$}
\label{app:alt_fit}

\begin{figure}[htbp]
\centering
\includegraphics[width=0.45\textwidth]{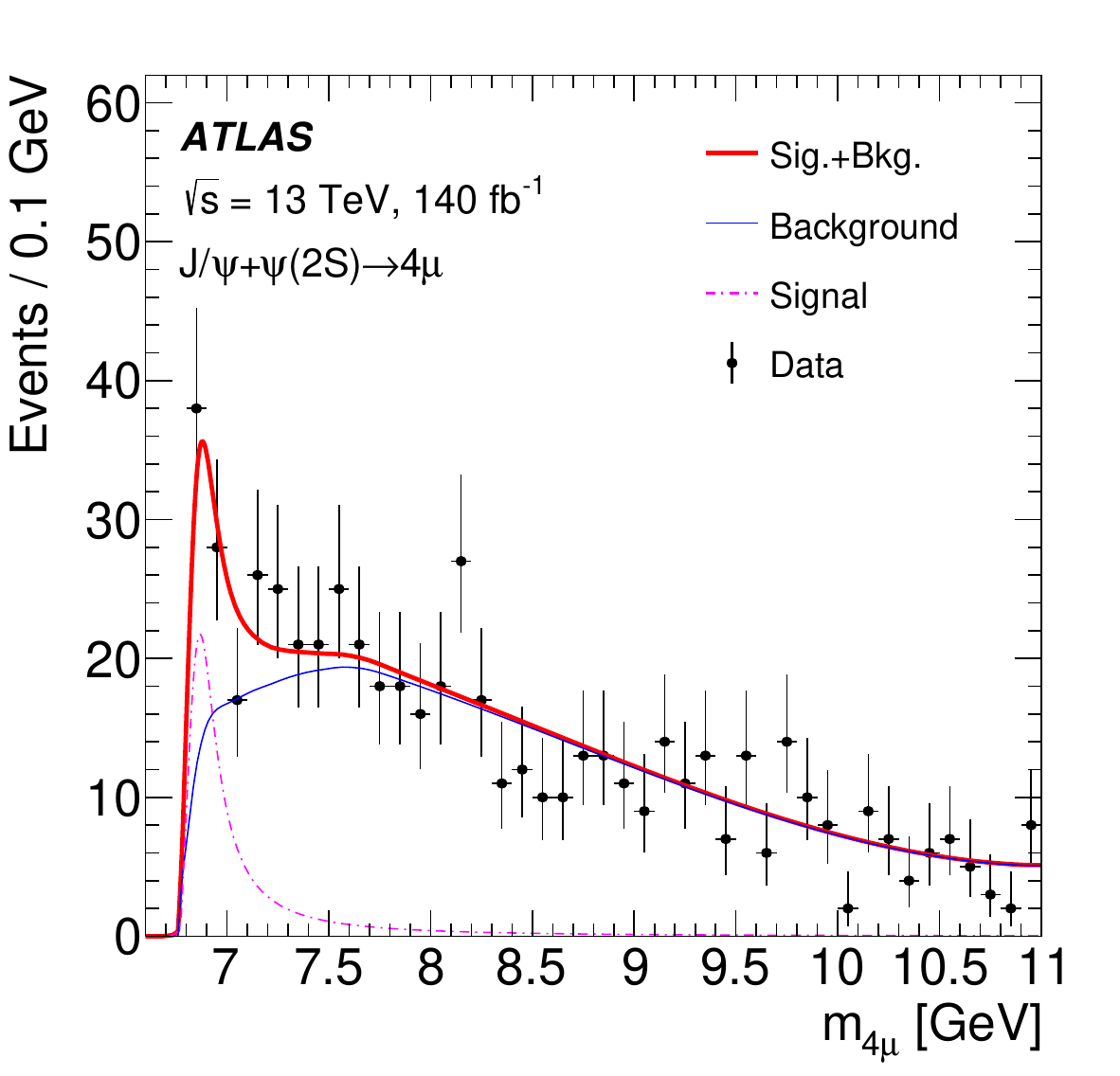}\\
\includegraphics[width=0.45\textwidth]{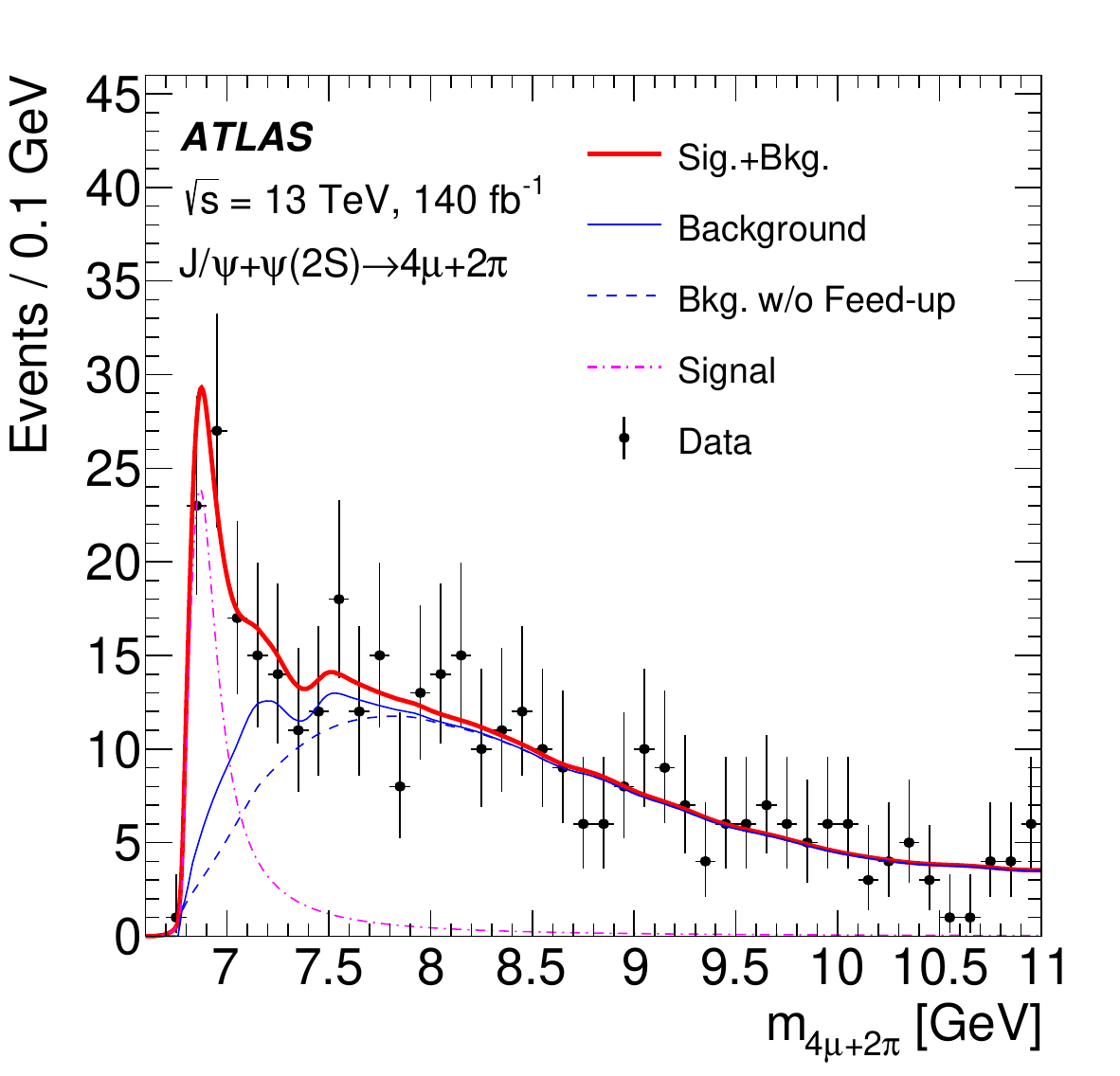}
\includegraphics[width=0.45\textwidth]{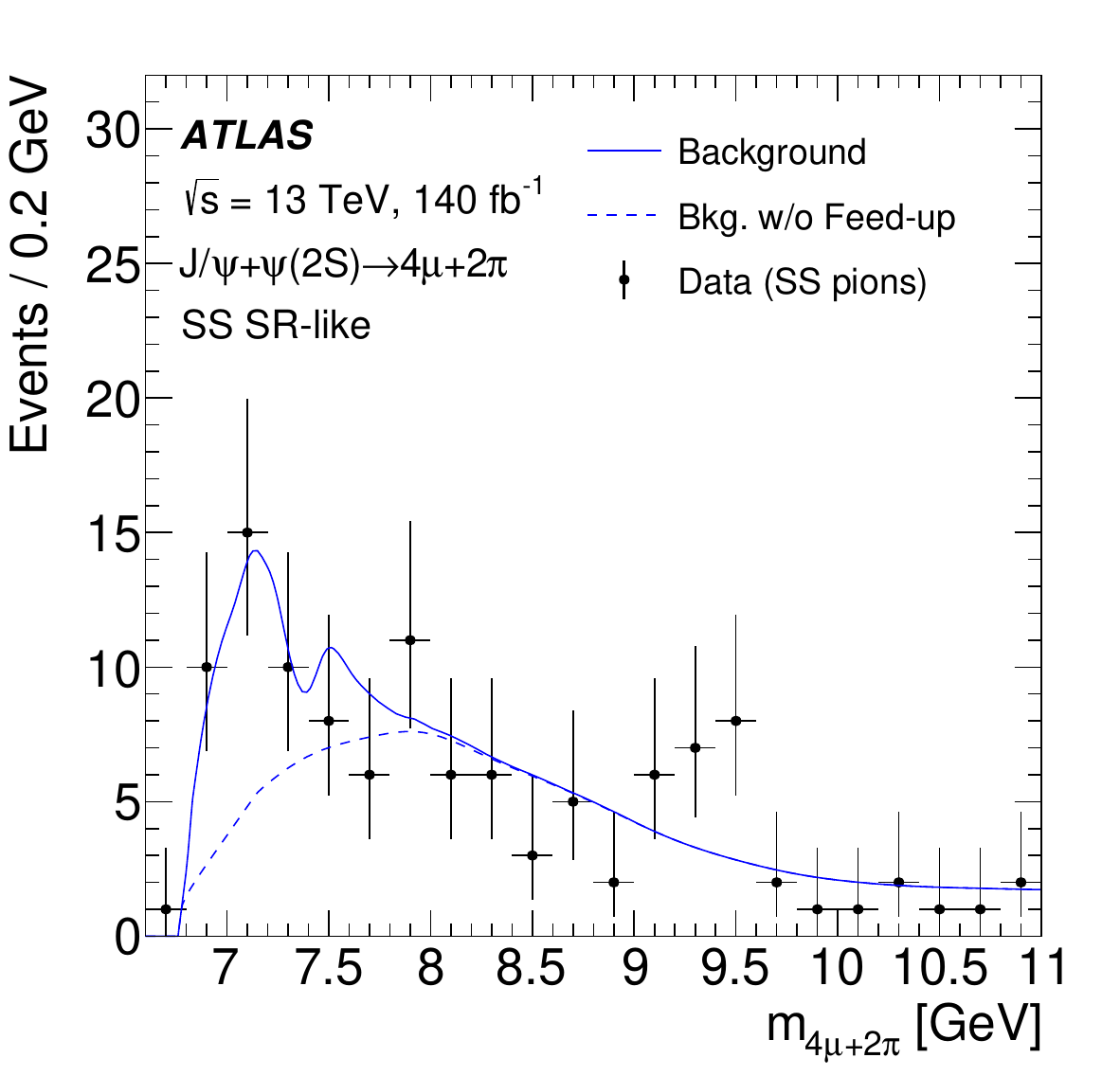}
\caption{Fits to the $J/\psi$+$\psi$(2S) mass spectra with model $C$ in the $4\mu$ (top), $4\mu+2\pi$ (bottom left), and $4\mu+2\pi$ SS SR-like region (bottom right), with SR and CR merged. The purple dash-dotted lines represent the signal resonances.}
\label{fig:modelC_nosplit}
\end{figure}

The spectra from an alternative fit using model $C$, in which the SRs and CRs are merged (i.e., without the $\Delta R$ splitting), are shown in Figure~\ref{fig:modelC_nosplit}. The signal remains clearly visible in both channel, and the fitted resonance mass and width are consistent with those from the nominal fit. The combined signal significance is 7.8$\sigma$ with the signal shape parameters fixed to their best-fit values, as in the nominal fit with the $\Delta R$ splitting.

\clearpage
\printbibliography
\clearpage
 
\begin{flushleft}
\hypersetup{urlcolor=black}
{\Large The ATLAS Collaboration}

\bigskip

\AtlasOrcid[0000-0002-6665-4934]{G.~Aad}$^\textrm{\scriptsize 104}$,
\AtlasOrcid[0000-0001-7616-1554]{E.~Aakvaag}$^\textrm{\scriptsize 17}$,
\AtlasOrcid[0000-0002-5888-2734]{B.~Abbott}$^\textrm{\scriptsize 122}$,
\AtlasOrcid[0000-0002-0287-5869]{S.~Abdelhameed}$^\textrm{\scriptsize 85b}$,
\AtlasOrcid[0000-0002-1002-1652]{K.~Abeling}$^\textrm{\scriptsize 55}$,
\AtlasOrcid[0000-0001-5763-2760]{N.J.~Abicht}$^\textrm{\scriptsize 49}$,
\AtlasOrcid[0000-0002-8496-9294]{S.H.~Abidi}$^\textrm{\scriptsize 30}$,
\AtlasOrcid[0009-0003-6578-220X]{M.~Aboelela}$^\textrm{\scriptsize 45}$,
\AtlasOrcid[0000-0002-9987-2292]{A.~Aboulhorma}$^\textrm{\scriptsize 36e}$,
\AtlasOrcid[0000-0001-5329-6640]{H.~Abramowicz}$^\textrm{\scriptsize 156}$
% ...
\bigskip
\\

% ...
$^{17}$Department for Physics and Technology, University of Bergen, Bergen; Norway.\\
% ...
$^{30}$Physics Department, Brookhaven National Laboratory, Upton NY; United States of America.\\
% ...
$^{36}$$^{(a)}$Facult\'e des Sciences Ain Chock, Universit\'e Hassan II de Casablanca;$^{(b)}$Facult\'{e} des Sciences, Universit\'{e} Ibn-Tofail, K\'{e}nitra;$^{(c)}$Facult\'e des Sciences Semlalia, Universit\'e Cadi Ayyad, LPHEA-Marrakech;$^{(d)}$LPMR, Facult\'e des Sciences, Universit\'e Mohamed Premier, Oujda;$^{(e)}$Facult\'e des sciences, Universit\'e Mohammed V, Rabat;$^{(f)}$Institute of Applied Physics, Mohammed VI Polytechnic University, Ben Guerir; Morocco.\\
% ...
$^{45}$Physics Department, Southern Methodist University, Dallas TX; United States of America.\\
% ...
$^{49}$Fakult\"{a}t Physik, Technische Universit{\"a}t Dortmund, Dortmund; Germany.\\
% ...
$^{55}$II. Physikalisches Institut, Georg-August-Universit\"{a}t G\"ottingen, G\"ottingen; Germany.\\
% ...
$^{85}$$^{(a)}$Khalifa University of Science and Technology, Abu Dhabi;$^{(b)}$New York University Abu Dhabi, Abu Dhabi;$^{(c)}$United Arab Emirates University, Al Ain;$^{(d)}$University of Sharjah, Sharjah; United Arab Emirates.\\
% ...
$^{104}$CPPM, Aix-Marseille Universit\'e, CNRS/IN2P3, Marseille; France.\\
% ...
$^{122}$Homer L. Dodge Department of Physics and Astronomy, University of Oklahoma, Norman OK; United States of America.\\
% ...
$^{156}$Raymond and Beverly Sackler School of Physics and Astronomy, Tel Aviv University, Tel Aviv; Israel.\\
% ...
$^{a}$ Also at Affiliated with an institute formerly covered by a cooperation agreement with CERN.\\
$^{b}$ Also at An-Najah National University, Nablus; Palestine.\\
$^{c}$ Also at Borough of Manhattan Community College, City University of New York, New York NY; United States of America.\\
$^{d}$ Also at Center for Interdisciplinary Research and Innovation (CIRI-AUTH), Thessaloniki; Greece.\\
$^{e}$ Also at Centre of Physics of the Universities of Minho and Porto (CF-UM-UP); Portugal.\\
$^{f}$ Also at CERN, Geneva; Switzerland.\\
% ...

\end{flushleft}


\end{document}